\documentclass[final,times,12p]{elsarticle}

\usepackage{amsmath}
\usepackage{amssymb}
\usepackage{enumerate}
\usepackage[margin=1.in]{geometry}
\usepackage{epsfig}
\usepackage{color}

\newcommand {\beq} {\begin{equation}}
\newcommand {\eeq} {\end{equation}}

\begin{document}

\begin{frontmatter}

\title{Stability of traveling waves in a driven Frenkel-Kontorova model}
\author[1]{Anna Vainchtein} \ead{aav4@pitt.edu}
\author[2,3]{Jes\'us Cuevas--Maraver \corref{cor1}} \ead{jcuevas@us.es}
\author[4,5]{Panayotis\ G.\ Kevrekidis} \ead{kevrekid@umass.edu}
\author[6]{Haitao Xu} \ead{hxumath@hust.edu.cn}

\address[1]{Department of Mathematics, University of Pittsburgh, Pittsburgh, PA 15260, USA}
\address[2]{Departamento de F\'{\i}sica Aplicada I, Escuela Politécnica Superior, Universidad de Sevilla, Sevilla 41011, Spain}
\address[3]{Instituto de Matem\'{a}ticas de la Universidad de Sevilla (IMUS). Edificio Celestino Mutis. Avda. Reina Mercedes s/n, 41012-Sevilla, Spain}
\address[4]{Department of Mathematics and Statistics, University of Massachusetts, Amherst, MA 01003-9305, USA}
\address[5]{Mathematical Institute, University of Oxford, OX26GG, UK}
\address[6]{Center for Mathematical Science, Huazhong University of Science and Technology, Wuhan, Hubei 430074, People's Republic of China}

\cortext[cor1]{Corresponding author}

\begin{abstract}
In this work we revisit a classical problem of traveling waves in a damped Frenkel-Kontorova lattice driven by a constant external force. We compute these solutions as fixed points of a nonlinear map and obtain the corresponding kinetic relation between the driving force and the velocity of the wave for different values of the damping coefficient. We show that the kinetic curve can become \emph{non-monotone} at small velocities, due to resonances with linear modes, and also at large velocities where the kinetic relation becomes \emph{multivalued}. Exploring the spectral stability of the obtained waveforms, we identify, at the level of numerical accuracy of our computations, a precise criterion for instability of the traveling wave solutions: monotonically decreasing portions of the kinetic curve always bear an unstable eigendirection. We discuss why the validity of this criterion in the {\it dissipative} setting is a rather remarkable feature offering connections to the Hamiltonian variant of the model and of lattice traveling waves more generally. Our stability results are corroborated by direct numerical simulations which also reveal the possible outcomes of dynamical instabilities.
\end{abstract}

\end{frontmatter}

\section{Introduction}
\label{sec:intro}

The Frenkel-Kontorova (FK) model was originally proposed in \cite{FK38} to describe dislocations in metals. The relevant setup involves the nonlinear dynamics of a chain of particles interacting with their nearest neighbors and a periodic onsite potential. Since its inception the model and its various extensions have been used to describe many physical phenomena, including dynamics of twin boundaries in crystals and domain walls in ferroelectric and ferromagnetic materials, crystal growth and surface reconstruction, charge-density waves, Josephson junctions in superconductors and DNA denaturation \cite{BraunKivshar04}. Both this discrete realization and the
corresponding continuum limit in the form of the (integrable) sine-Gordon equation
have constituted a basis for extensive studies that have by now been summarized
in various books~\cite{BraunKivshar04,cuevasbook}. These works include detailed investigations of traveling wave solutions in the Hamiltonian FK model both in the presence \cite{PK84} and in the absence \cite{Aigner03} of the external drive.

Given its apparently non-integrable nature,
much of the analytical progress in understanding the dynamics of driven FK chains, in which each particle is subjected to a constant force, has been limited to models with piecewise quadratic continuous onsite potentials emulating the onsite nonlinearity of the original FK model. As first shown by Atkinson and Cabrera in \cite{AC65}, if each well is represented by a convex parabola, one can use Fourier transform techniques to derive an exact analytical solution of the system's dynamics in the form of a traveling wave, also known as a (generalized) \emph{kink}. The wave connects equilibria in two neighboring wells and features short-length oscillations emitted by the moving front. A more detailed analysis of this solution and its extension to the case with damping can be found in \cite{Carpio03,EW77,Kresse02,KT03,KT07}.
The resulting traveling wave shares many features with the semi-analytical solution constructed in \cite{CF70,Ishioka71} for a higher-dimensional version of the model. The formal Atkinson-Cabrera solution for the conservative problem yields a force-velocity diagram that has multiple \emph{resonances} at low speeds that coincide with group velocities of emitted waves. Further analysis, however, has revealed that the solution does not satisfy the inequalities used to obtain it, and hence is not admissible, below a certain threshold velocity \cite{KT03,EW77} that includes all of the resonances, and the same is true for the solution in \cite{CF70,Ishioka71}.
Using the semi-analytical approach developed in \cite{FCC77} for a higher-dimensional version of the problem, traveling wave solutions for the conservative FK problem with smoother piecewise quadratic models of the onsite potential, where two convex parabolas are connected by a concave one, were constructed in \cite{V10}. In \cite{RV13} the approach was adopted to fill in the low-velocity gap left by the Atkinson-Cabrera solutions (see also \cite{LVW14} for the analogous results in the higher-dimensional case). These results reveal highly non-monotone kinetic relations $\sigma(c)$ between the applied force $\sigma$ and velocity $c$ of the traveling wave, with cusps at the resonance velocities. Parts of these curves at medium to high velocity values were also obtained in \cite{EW77} using different methods. Results of numerical simulations in \cite{EW77,RV13,V10} strongly suggest that $\sigma'(c)>0$ is necessary but not sufficient for stability of the obtained solutions, and the conjecture apparently also applies to the underdamped case \cite{Carpio04,RV13}.
A complementary perspective of this criterion is that decreasing portions
of the force-velocity curve (with $\sigma'(c)<0$) are guaranteed to lead
to at least one unstable eigendirection and instability of the corresponding traveling
waves.
A heuristic argument for the necessity part of this conjecture was proposed in \cite{EW77}. One can physically think of the relevant argument as follows: a small positive perturbation of velocity accelerates the wave because the driving force is larger than the one necessary to
keep the new velocity constant~\cite{Strunz98}.
For the underdamped FK problem with a two-parabola onsite potential, stability of sufficiently fast waves was proved in \cite{Carpio04} but for technical reasons the result does not extend to the last minimum of the kinetic curve, as originally conjectured in \cite{AC65}, and the proof thus does not connect the stability threshold to the sign change of $\sigma'(c)$. Loss of stability via Hopf bifurcations as some parameters are varied in a related model with tilted piecewise quadratic and quartic potentials was numerically demonstrated in \cite{Shiroky18}, with analytical results obtained for the two-parabola case.

In the present work we consider the stability of traveling waves in the underdamped driven FK problem with the original sinusoidal nonlinearity, thus focusing on the fully nonlinear case free of the degeneracies of the piecewise linear approximations.
While the conclusions of the latter are particularly
insightful, our study is motivated by the numerical results and discussion in \cite{PK84} for the conservative dynamics of the original FK model at small driving force, which provide strong evidence that some of the resonances observed in the piecewise linear problem also play an important role in the fully nonlinear case. To fully understand these results, which were obtained by conducting direct numerical simulations at given force $\sigma$ and thus could only capture stable solutions within the basin of attraction of the initial data, it is necessary to compute the traveling wave solutions and obtain the full kinetic curve as in \cite{V10}. However, due to the non-decaying quasiperiodic tail oscillations, computing traveling waves in the conservative problem has proved to be a particularly difficult task with the existing numerical techniques. When viscosity is included, choosing a long enough chain allows the wave to approach the equilibrium states at the ends within the numerical error. The length of the chain necessary to obtain accurate results increases as the damping coefficient becomes smaller. Setting the number of particles to be large enough, we obtain traveling wave solutions for small and intermediate values of the damping coefficient.

Our computational approach is based on the observation that traveling wave solutions are periodic modulo a shift by one lattice site and
thus are fixed points of the corresponding (propagate by one site and shift back) nonlinear map. { We use a numerical procedure based on Newton-Raphson iterations to compute these fixed points}. Focusing on the low-velocity regime at smaller damping, we compute the corresponding kinetic curve $\sigma(c)$ that features non-monotone behavior around the same resonance velocities as in \cite{PK84}. We then investigate the stability of the obtained solutions by deploying the approach used in our recent work \cite{jxu1,jxu2} and computing the Floquet multipliers of the corresponding monodromy matrix for the relevant map that takes into account periodicity modulo lattice shift.
The results show that the decreasing portions of the kinetic curve indeed correspond to unstable traveling waves, while the other waves appear to be linearly stable within the accuracy of the computations. The transition between unstable and stable states takes place when a real Floquet multiplier crosses the unit circle on the right.
Stability of at least some solutions along the increasing parts of the kinetic curve is further confirmed by direct numerical simulations with generic initial data and by dynamic evolution runs generated by perturbing the unstable waves along the corresponding eigenmode, both resulting in relaxation to a stable traveling wave pattern.

Traveling wave solutions yielding non-monotone kinetic curves with multiple resonances were computed in the earlier work on damped driven FK problem but these studies focused on the ring configuration with periodic boundary conditions (modulo $2\pi M$, where $M$ is the number of kinks trapped in the ring and $2\pi$ is the distance between two neighboring wells in the periodic onsite potential) with small to medium number of particles \cite{Abell05,Strunz98,Watanabe96}, and Floquet computations were
performed in \cite{Strunz98} and \cite{Watanabe96} to investigate stability. However, the obtained solutions are generally quite different from the ones considered here in that they do not relax to equilibrium states at the boundaries for all velocities and either involve a very small (four to eight) number of particles \cite{Watanabe96} or contain multiple kinks trapped in the ring \cite{Strunz98}.
This modifies the values of the resonance velocities and generally results in different kinetics at large speeds \cite{Strunz98,Watanabe96}. Nevertheless, despite
the significant deviations between the settings, these authors too observe instability of their solutions along the decreasing portions of the kinetic curves. This suggests the potentially broader
relevance of the corresponding observation.
Further numerical evidence and discussion of resonances in this problem can be found in \cite{Ishioka73,Ustinov93,Zheng98}.

At larger viscosity coefficient, the peaks due to resonances disappear, and the kinetic curve becomes monotone in the small-velocity regime. However, as we show, at large enough velocities the kinetic relation $\sigma(c)$ loses monotonicity and in fact becomes \emph{multivalued}, with the primary branch reaching a maximum at a certain critical force $\hat{\sigma}_1$ above which the $2\pi$-kink traveling wave solutions no longer exist. At velocities above $\hat{c}_1$ corresponding to this critical value the force along the primary branch of the kinetic relation decreases until the curve reaches a turning point, giving rise to another branch of the kinetic relation. The critical force $\hat{\sigma}_1$ has been identified in the earlier work \cite{Braun00}, which also investigated the mechanism and the consequences of the stability loss, but since that study relied on a numerical continuation in the driving force that enabled computation of only stable attractors, it missed the fact that the kinetic curve in fact continues beyond the corresponding velocity $\hat{c}_1$ and then turns around. The turning point corresponds to the \emph{maximal velocity} of the traveling wave. Moreover, the new branch of the kinetic relation emanating from this point eventually reaches a minimum point and continues on until another turning point giving rise to yet another branch, and so on, { with the kinetic curve proceeding in a spiral fashion. While such spiraling kinetic relations are known in the continuum setting of damped driven sine-Gordon equation \cite{Brown94,Maksimov96,vdBerg03}, to our knowledge, they have not been previously observed for the discrete FK problem.} Floquet analysis shows that all traveling wave solutions beyond the critical point $(\hat{c}_1,\hat{\sigma}_1)$ on the kinetic curve are unstable. We note that this does not contradict the stability criterion which states that $\sigma'(c)<0$ is sufficient but not necessary for instability. Importantly, every extremal point along the multivalued kinetic curve is associated with the emergence of a new unstable eigendirection corresponding to another real multiplier crossing the unit circle on the right as $\sigma(c)$ decreases.

The paper is organized as follows. In Sec.~\ref{sec:TW} we formulate the problem and discuss some general features of the traveling wave solutions. Numerical methods and results are discussed in Sec.~\ref{sec:numer}. We discuss the results and identify future research directions in Sec.~\ref{sec:conclusions}.

\section{Traveling waves in a damped driven Frenkel-Kontorova model}
\label{sec:TW}

Consider an infinite chain of particles of mass $m$ connected by linear springs of stiffness $K>0$ to their nearest neighbors and
interacting with an external periodic substrate potential $\Psi(u_n)$ with multiple wells, where $u_n(t)$ is the vertical displacement of the $n$th mass from its reference position at time $t$. We assume that the particles can move only along the vertical
direction and that a constant force $\sigma$, acting in the same direction, is applied to each particle. The substrate potential has period $a$ and satisfies
\[
\Psi'(0)=\Psi'(a)=0, \quad \Psi''(0)=\Psi''(a)=G>0.
\]
The equations of motion are
\[
m \ddot{u}_n+\eta \dot{u}_n=K(u_{n+1}-2u_n+u_{n-1})+\sigma-\Psi'(u_n),
\]
where dots denote time derivatives, and we have included damping with the coefficient $\eta \geq 0$. Introducing the rescaled quantities
\[
\bar{t}=\dfrac{t\sqrt{K}}{\sqrt{m}}, \quad \bar{u}_n=\dfrac{2\pi}{a} u_n, \quad \bar{\sigma}=\dfrac{2\pi}{G a}\sigma, \quad
\bar{\Psi}=\dfrac{(2\pi)^2}{G a^2}\Psi
\]
and dropping the bars, we obtain
\beq
\ddot{u}_n+\gamma \dot{u}_n=u_{n+1}-2u_n+u_{n-1}+\mu(\sigma-\sin u_n),
\label{eq:dyn}
\eeq
where
\beq
\gamma=\dfrac{\eta}{\sqrt{K m}}, \qquad \mu=\dfrac{G}{K}
\eeq
are the dimensionless parameters measuring viscosity and relative strength of the nonlinear interactions, respectively, and we set
the rescaled $2\pi$-periodic potential to $\Psi(u)=1-\cos(u)$, as in the original Frenkel-Kontorova model \cite{BraunKivshar04,FK38}.

We seek traveling wave solutions of \eqref{eq:dyn} in the form
\beq
u_n(t)=\phi(\xi), \quad \xi=n-ct,
\label{eq:TWansatz}
\eeq
where $c>0$ is the velocity of the wave. These solutions satisfy the advance-delay differential equation
\beq
c^2 \phi''(\xi)-\gamma c \phi'(\xi)=\phi(\xi+1)-2\phi(\xi)+\phi(\xi-1)+\mu(\sigma-\sin(\phi(\xi)))
\label{eq:TWeqn}
\eeq
and connect the equilibrium states in the neighboring wells:
\beq
\phi(\xi) \to \text{arcsin}(\sigma) + \begin{cases} 2\pi, & \xi \to -\infty\\
                                                         0, & \xi \to \infty,
                                     \end{cases}
\label{eq:BCs}
\eeq
{ where we assume $0 \leq \sigma < 1$}.
As we will illustrate in the Sec.~\ref{sec:numer}, the traveling wave solutions only exist when $c$ and $\sigma$ satisfy a particular kinetic relation. Moreover, due to the translational invariance of \eqref{eq:TWeqn}, such solutions (when they exist) are not unique for a given $(c,\sigma)$ pair. To select a unique solution, it suffices to impose a
pinning condition, e.g.
\beq
\phi(0)=\pi.
\label{eq:pin}
\eeq

\begin{figure}
\centerline{\psfig{figure=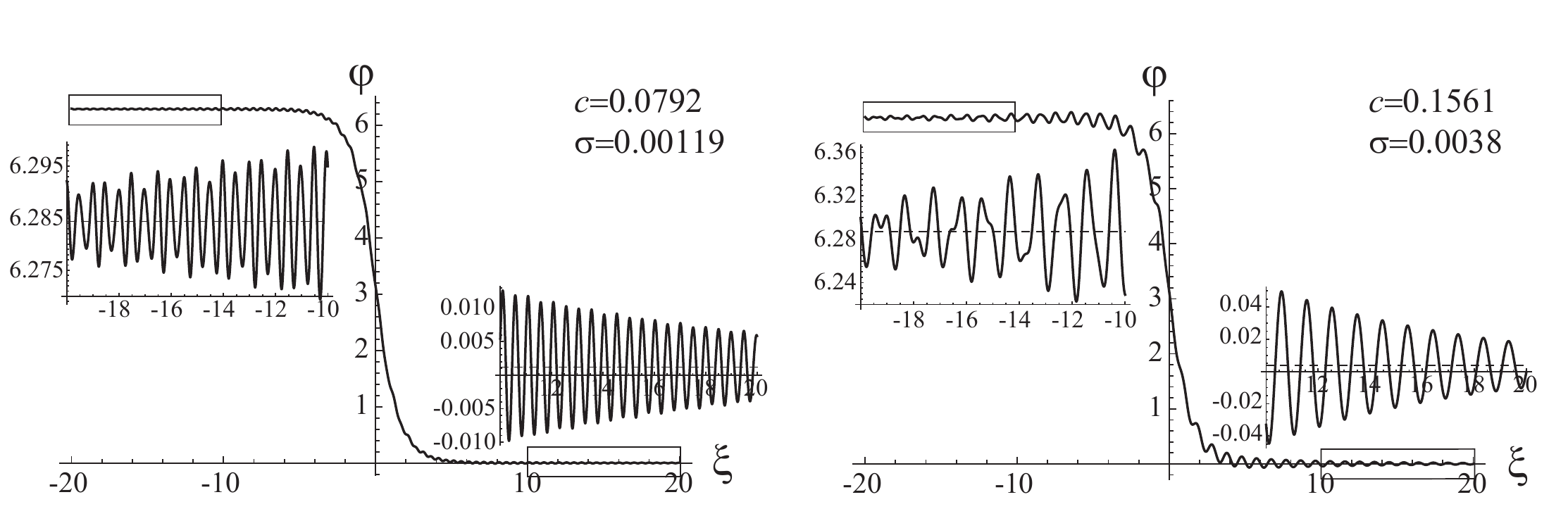,width=\textwidth}}
\caption{\footnotesize Examples of traveling waves and their oscillatory tails obtained from numerical simulations, as described in Sec.~\ref{sec:numer}. Insets zoom in on the tail oscillations. Here $\mu=1$ and $\gamma=0.01$.}
\label{fig:TWexamples}
\end{figure}

The traveling wave solutions feature oscillations in one or both tails that decay at infinity, as illustrated in Fig.~\ref{fig:TWexamples}. %
To obtain their asymptotic form, it suffices to linearize \eqref{eq:TWeqn} about either equilibrium state (both lead to the same result) and seek a solution of the resulting linear problem in the plane-wave form $\exp(ik\xi)$. This yields the characteristic equation
\beq
L(k) \equiv 4\sin^2(k/2)-i\gamma c k +\mu\cos(\text{arcsin} \ \sigma)-c^2 k^2=0
\label{eq:L}
\eeq
for the wave numbers $k$ of the tail oscillations in a traveling wave corresponding to $(c,\sigma)$ pair. The fact that the solutions decay at infinity means that modes with wave numbers satisfying $\text{Im} k>0$ appear ahead of the moving front, while modes with $\text{Im} k<0$ appear behind. The structure of the roots of \eqref{eq:L} was studied in detail in \cite{Carpio03,Kresse02,KT03,KT07}. When $\gamma>0$, there are infinitely many complex roots that all have $\text{Im} k \neq 0$. But in the absence of viscosity there is a finite number of real roots in addition to infinitely many roots with nonzero imaginary part. The real roots are nonzero and symmetric about the imaginary axis. They correspond to non-decaying tail oscillations associated with the radiation emitted by the traveling wave front as it moves through the lattice. The real-$k$ modes are distributed according to the radiation condition as follows. Modes with phase velocity $c=\omega(k)/k$
(where the frequency $\omega$ satisfies the dispersion relation $\omega^2(k)=4\sin^2(k/2)+\mu\cos(\text{arcsin} \ \sigma)$) below the group velocity, $c<c_g$, where $c_g=\omega'(k)$, propagate behind the traveling wave front and modes with $c>c_g$ propagate ahead of it. Velocity values that coincide with group velocity for some mode correspond to \emph{resonances}. At non-resonance $c$, the number of radiation modes propagating behind always differs by one from the number of modes ahead
and the total (odd) number of modes changes as the resonance velocities are crossed \cite{KT03,PK84}. In particular, at velocities above the first resonance value, there is a single radiation mode behind the front and no modes ahead, so that the solution decays monotonically to zero as $\xi \to \infty$. As shown in \cite{Carpio03,Kresse02}, when a small damping is introduced ($0<\gamma \ll 1$), the real roots move into respective upper and lower half-planes in agreement with the radiation condition, with $|\text{Im} k|$ proportional to $\gamma c$. This means that at small $\gamma c$ there are oscillations that decay very slowly at infinity, making it necessary to consider chains in numerical computations of traveling waves that are sufficiently long to ensure that the conditions \eqref{eq:BCs} at infinity are well approximated.

Observe also that the lattice traveling waves $u_n(t)=\phi(\xi)$ are periodic modulo shift by one lattice space. Indeed, \eqref{eq:TWansatz} implies that
\beq
u_{n+1}(t+T)=u_n(t),
\label{eq:per_shift}
\eeq
where $T=1/c$ is the period. Hence the traveling wave solutions can be cast as fixed points of the nonlinear map
\begin{equation}\label{eq:nonlin_map}
\left[\begin{array}{c}
  \{u_{n+1}(T)\} \\ \{\dot u_{n+1}(T)\} \\  \end{array}\right]
  \rightarrow
  \left[\begin{array}{c}
  \{u_{n}(0)\} \\ \{\dot u_{n}(0)\} \\  \end{array}\right],
\end{equation}
which consists of solving the governing equations \eqref{eq:dyn} over one period and then shifting the result back by one lattice space.
In Sec.~\ref{sec:numer}, we will use this representation to compute the traveling waves numerically.
Importantly, identifying the traveling waves as fixed points of this map will enable us to use
Floquet theory to examine their spectral stability.

\section{Numerical results}
\label{sec:numer}

\subsection{Numerical methods for computing traveling waves and stability analysis}
\label{sec:methods}

In order to obtain a traveling wave, we employ a fixed point method { similar to the one used for calculating discrete breathers described e.g. in
  the prototypical work of~\cite{MarinAubry96} for Hamiltonian nonlinear lattices}. Considering a lattice with $N$ particles, where $N$ is even,
we use { a Newton-Rapshon method for solving the map defined in (\ref{eq:nonlin_map})}. In other words, we apply the map $\mathbf{Y}\rightarrow\mathbf{Y}-\mathbf{J}^{-1}\mathbf{F}$ where $\mathbf{J}$ is the finite-difference Jacobian matrix of the nonlinear map $\mathbf{F}(\mathbf{Y})$ defined as
\begin{equation}\label{eq:fshoot}
\mathbf{F}(\mathbf{Y})=\left[\begin{array}{c}
  \{u_{n+1}(T)-u_{n}(0)\} \\ \{\dot u_{n+1}(T)-\dot u_{n}(0)\} \\ u_0(T)-\pi \end{array}\right], \quad \mathbf{Y}=\left[\begin{array}{c}
  \{u_{n}(0)\} \\ \{{\dot u}_{n}(0)\} \\ \alpha \end{array}\right], \quad n=-N/2, \ldots, N/2-1
\end{equation}
with $\alpha$ in the vector of variables $\mathbf{Y}$ being either the force $\sigma$ or the velocity $c$, depending on whether the parameter continuation is performed in $c$ or $\sigma$, respectively. The last row in $\mathbf{F}(\mathbf{Y})$ corresponds to the pinning condition (\ref{eq:pin}). In addition, one must take into account the boundary conditions to define $u_{N/2}$ and $\dot{u}_{N/2}$ in \eqref{eq:fshoot}. For example, if the periodic (modulo $2 \pi$ in the field $u$) boundary conditions
\beq
u_{-N/2-1}=u_{N/2-1}+2\pi, \quad u_{N/2}=u_{-N/2}-2\pi
\label{eq:per_BCs}
\eeq
are implemented, one should use $u_{-N/2}-2\pi$ and $\dot u_{-N/2}$ instead of $u_{N/2}$
and $\dot u_{N/2}$, respectively, in $\mathbf{F}(\mathbf{Y})$. In some computations, we use the free end boundary conditions
\beq
u_{-N/2-1}=u_{-N/2}, \quad u_{N/2}=u_{N/2-1}
\label{eq:free_BCs}
\eeq
instead. However, for large values of $N$ used in our work the specific choice of boundary conditions is not essential since the obtained traveling waves approach the equilibrium values at the boundaries, and thus the solutions computed, for example, using \eqref{eq:free_BCs} also satisfy \eqref{eq:per_BCs} within their numerical accuracy.
In order to get the values of the solution at $t=T$ we integrate the dynamical equations $(\ref{eq:dyn})$ by means of the Dormand-Prince algorithm (Matlab's \texttt{ode45} function).

To investigate the linear stability of traveling waves, we substitute $u_n(t)=v_n(t)+\epsilon \xi_n(t)$ into (\ref{eq:dyn}) and consider $O(\epsilon)$ terms resulting from this perturbation, yielding the following equation:
\beq
\ddot{\xi}_n+\gamma \dot{\xi}_n=\xi_{n+1}-2\xi_n+\xi_{n-1}-(\mu\cos u_n)\xi_n,
\label{eq:perturb}
\eeq

The Floquet multipliers $\lambda$ are the eigenvalues of the monodromy matrix $\mathcal{M}$ defined by the map
\begin{equation}\label{eq:monodromy}
\left[\begin{array}{c}
  \{\xi_{n+1}(T)\} \\ \{\dot \xi_{n+1}(T)\} \\  \end{array}\right]
  =\mathcal{M}
  \left[\begin{array}{c}
  \{\xi_{n}(0)\} \\ \{\dot \xi_{n}(0)\} \\  \end{array}\right].
\end{equation}
The Dormand-Prince algorithm is again used to integrate (\ref{eq:perturb}).

\subsection{Instability due to low-velocity resonances at small damping}
\label{sec:small_gamma}
Using the method described in Sec.~\ref{sec:methods} with free end boundary conditions \eqref{eq:free_BCs}
and $N=8000$ particles, we have computed traveling wave solutions for given velocities at $\gamma=0.01$ and $\mu=1$.
In what follows, we show results for the velocity interval $0.035 \leq c \leq 0.254$.
Slow decay of large-amplitude tail waves at higher velocities makes it necessary to consider
progressively longer chains to avoid spurious oscillations in the numerical procedure, and resolving the piled-up resonances at small enough velocities (as discussed below) is also quite
computationally intensive.

For comparison we also conducted direct numerical simulations at given $\sigma$ with fixed end boundary conditions 
$u_0=\text{arcsin}(\sigma)+2\pi$, $u_L=\text{arcsin}(\sigma)$ and initial conditions $u_n(0)=u_n^S(0)+\text{arcsin}(\sigma)$, $\dot{u}_n(0)=0$, where $u_n^S$ is a stable static kink at zero force (computed by solving the problem at $\sigma=0$ with viscosity $\gamma=1$, piecewise constant initial displacement and zero initial particle velocity).
The number $L$ of particles and the simulation time $t_\text{max}$ were chosen large enough to allow steady motion to develop without being affected by wave reflections from the boundaries. In a typical small-velocity simulation, we set $L=1600$ and $t_\text{max}=3000$. To compute the velocity $c$ of the resulting traveling wave at given $\sigma$, we found the times $t_n$ such that $u_n(t_n)=\pi$ for a range of particles crossing from one well to another during the course of the simulation and computed the instantaneous velocities $c_n=1/(t_{n+1}-t_n)$, which converged to a constant value at large $n$. The wave velocity $c$ was then determined as the average of $c_n$ over the last twenty time periods.

\begin{figure}
\centerline{\psfig{figure=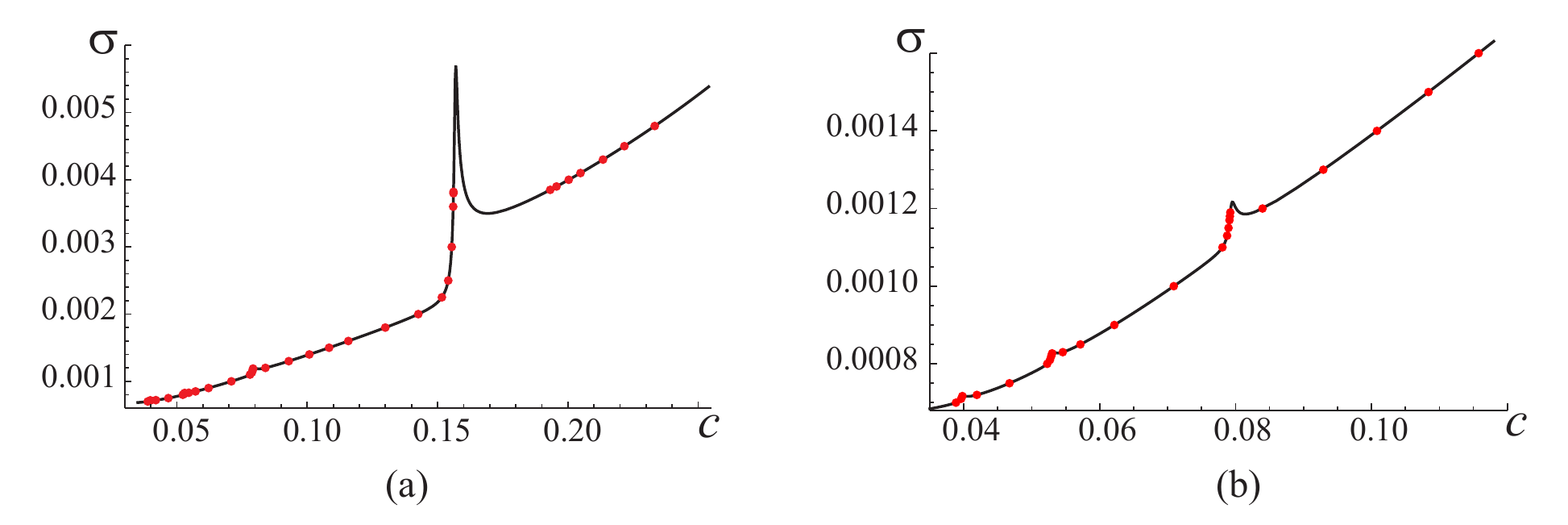,width=\textwidth}}
\caption{\footnotesize (a) Force as a function of velocity (solid curve) for the traveling wave solutions and the results of direct numerical simulations (dots). (b) Zoomed-in version at smaller velocities. Here $\mu=1$ and $\gamma=0.01$.}
\label{fig:kinetics_gam0p01}
\end{figure}

The force-velocity diagram (solid curve) resulting from the { fixed-point method}, as well as the direct numerical results (dots) are shown in Fig.~\ref{fig:kinetics_gam0p01}.
One can see an excellent agreement between the kinetic curve obtained by the numerical
fixed point method and the direct simulations. Note, however, that while a numerical simulation can only capture a stable solution within the basin of attraction of the initial data, the  former
fixed point procedure allows us to compute the entire $\sigma=\sigma(c)$ curve, including the velocity intervals that correspond to unstable solutions, as discussed below.
The kinetic curve $\sigma(c)$ is non-monotone around certain resonance velocities, with $\sigma$ rapidly increasing over a narrow velocity interval, reaching a local maximum and then decreasing more gradually until it reaches a local minimum. The amplitude of these resonance maxima
is larger at higher velocities. The rapidly increasing portions are reminiscent of the numerical results in \cite{PK84} for the undamped driven problem, which revealed a step-like dependence of $c$ on $\sigma$ (prescribed in the simulations), with $\sigma$ increasing at nearly constant $c$ over each step (see Fig.~6 in \cite{PK84}, where $\mu=1$ as in our Fig.~\ref{fig:kinetics_gam0p01} but $\gamma=0$). In fact, our direct numerical simulations yield similar results (marked by dots), with shorter $\sigma$ intervals over each ``step'' and larger values of $\sigma$ due to nonzero damping. This is particularly evident for the largest resonance, where $\sigma(c)$ rapidly grows for $0.154 \leq c \leq 0.157$, though the numerical simulation results only capture a portion of this ``step'' before switching over to larger $c$. More generally, it is clear that these results do not reveal the complete picture around the resonances that can be obtained from the $\sigma(c)$ curve. Nevertheless, we observe that none of the solutions obtained via direct numerical simulations correspond
to the decreasing portions of the $\sigma(c)$ curve, which already suggests instability of the waves with $\sigma'(c)<0$.

\begin{figure}
\centerline{\psfig{figure=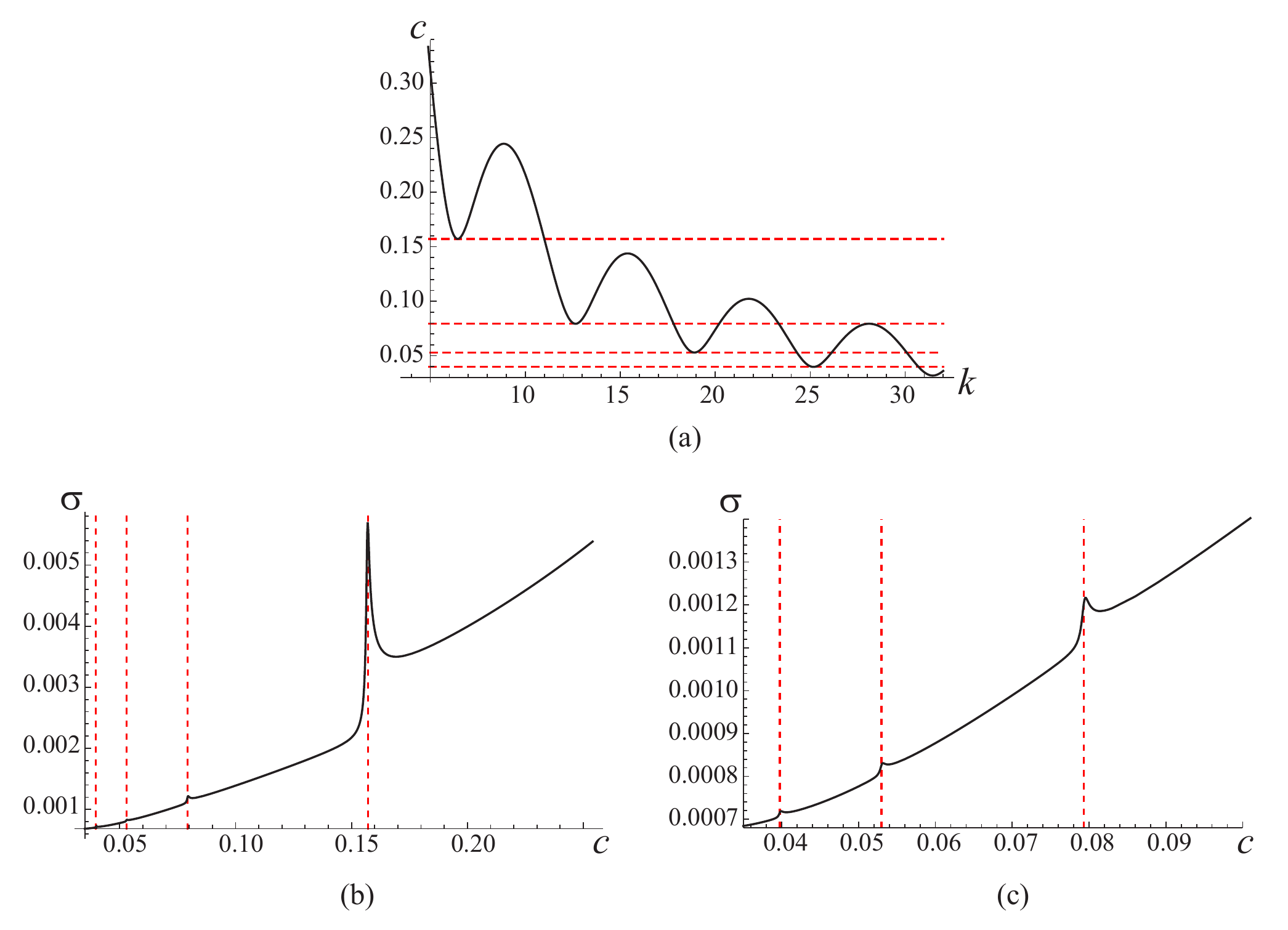,width=\textwidth}}
\caption{\footnotesize (a) Velocity $c$ as a function of the real wave number $k$ of the linear waves in the chain with no damping ($\gamma=0$) and zero force (solid curve). The dashed horizontal lines mark the velocity values corresponding to the first four minima. Here $\mu=1$. (b) The kinetic curve $\sigma(c)$ (solid curve) at $\gamma=0.01$, $\mu=1$ shown together with the velocity values in (a) (dashed vertical lines). (c) Enlarged view of the small-velocity region.}
\label{fig:res_velocities}
\end{figure}

To obtain approximate values of the velocities corresponding to the local maxima of $\sigma(c)$, we follow \cite{PK84} (see also related discussions in \cite{Strunz98,Ustinov93,Watanabe96,Zheng98}) and consider again the linearized problem discussed in Sec.~\ref{sec:TW}. Observe that both $\gamma$ and $\sigma$ are small, so the corresponding contributions to the characteristic equation \eqref{eq:L} for the wave numbers $k$ of the tail oscillations can, to the leading order, be neglected. One then obtains
\[
4\sin^2(k/2)+\mu-c^2 k^2=0,
\]
which can be solved for $c$ as a function of real wave numbers $k$ corresponding to radiated waves in the conservative problem. The result is shown by a solid curve in Fig.~\ref{fig:res_velocities}(a).
The curve has an infinite number of minima and maxima corresponding to resonance velocities mentioned above in Sec.~\ref{sec:TW}.
The intersection of the curve with a horizontal line corresponding to some non-resonance velocity $c$ yields a finite odd number of positive wave numbers corresponding to emitted lattice waves. According to the radiation condition (or, equivalently, zero-viscosity limit, as mentioned above), the modes along the decreasing portions of the curve appear behind the traveling front and modes along the increasing parts appear ahead of it \cite{Carpio03,KT03}. At the velocity values corresponding to the \emph{local minima} of the $c(k)$ curve, two modes, one propagating ahead and one behind, merge, so that the total number of modes decreases by two as $c$ crosses this value from above. Similarly, the number of modes increases by two as the local maxima of $c(k)$ curve are crossed from above. The first four velocities corresponding to the local minima, $c_1^* \approx 0.1572$, $c_2^* \approx 0.0793$, $c_3^* \approx  0.0530$ and $c_4^* \approx 0.0398$, are marked by the dashed horizontal lines lines in the figure. As shown in Fig.~\ref{fig:res_velocities}(b) and Fig.~\ref{fig:res_velocities}(c), these resonance velocities (marked by dashed vertical lines), obtained from the linear theory that neglects damping and nonzero $\sigma$, are very close to the values of $c$ corresponding to the local maxima of $\sigma(c)$. The same velocity values were used to explain the numerical results in \cite{PK84}.

It is unclear why the system does not exhibit resonances at the values close to the local maxima of $c(k)$. Some insight into this may be gleamed from the problem with a double-well onsite potential represented by two convex parabolas that are connected by a concave one. As shown in \cite{V10}, where a driven undamped problem with such potentials was studied, when the width $\delta$ of the concave region is sufficiently small, the curve $\sigma(c)$ exhibits resonances (represented by cusps) at both minima and maxima of $c(k)$. However, as $\delta$ increases while the distance between the two wells is kept fixed, only the resonances at the local minima remain, and they acquire the structure similar to that seen in Fig.~\ref{fig:kinetics_gam0p01}. Thus it appears that a sufficiently wide concave region within the onsite potential (relative to the distance between the two wells) plays a role in suppressing the resonances at the local maxima of $c(k)$.

\begin{figure}
\centerline{\psfig{figure=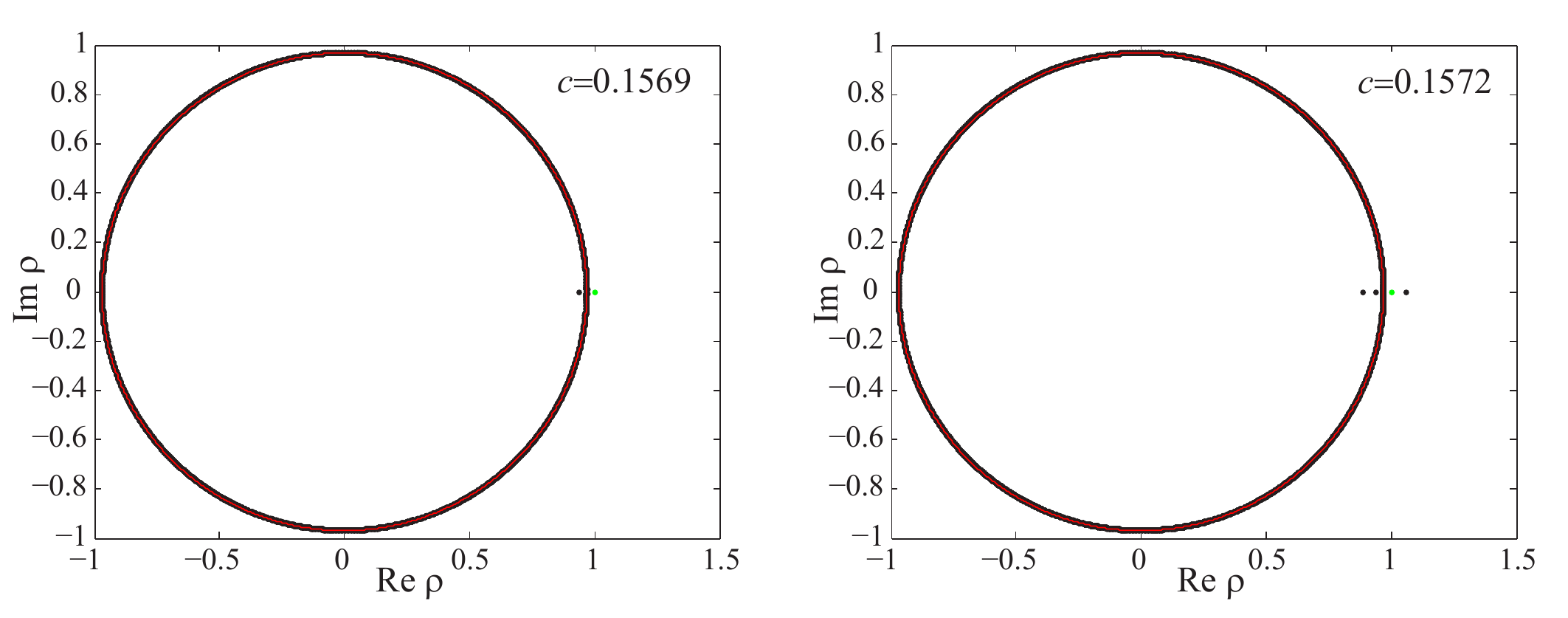,width=\textwidth}}
\caption{\footnotesize Floquet multipliers (dots) for traveling wave solutions at $c=0.1569$ (left panel) and $c=0.1572$ (right panel). The red curve marks the circle of $|\rho|=\exp[-\gamma/(2c)]$. Here $\mu=1$, $\gamma=0.01$, { and the multiplier $\rho=1$ is marked by a green dot}.}
\label{fig:Floquet_examples}
\end{figure}

We now consider the stability of the obtained solutions. Using the procedure described in Sec.~\ref{sec:methods}, we computed the Floquet multipliers $\rho$ for each of the obtained traveling wave solutions. One can show \cite{Watanabe96} that in the problem with damping, $\rho$ either lie on the circle $|\rho|=\exp[-\gamma/(2c)]$ or come in pairs $\rho$, $\hat{\rho}$ such that $\rho \hat{\rho}=\exp(-\gamma/c)$. They are either real or come in complex conjugate pairs. In particular, there is always a real multiplier $\rho_1=1$ corresponding to the time-translational invariance and its real counterpart $\hat{\rho}_1=\exp(-\gamma/c)$ \cite{Braun00}.
Examples of computed Floquet mutipliers are shown in Fig.~\ref{fig:Floquet_examples} by dots, with the red curves marking the circle $|\rho|=\exp[-\gamma/(2c)]$ in each case.
We note (for purposes of the discussion below) that for the typical parameter values shown in the
figure, the relevant real multipliers can be clearly discerned to be separated from the eigenvalue
circle of radius $|\rho|=\exp[-\gamma/(2c)]$.
In the left panel ($c=0.1569$) there are two real multipliers, $\rho_1=1$ and $\hat{\rho}_1=\exp(-\gamma/c)$ (within the numerical accuracy). Most of the non-real multipliers lie on the circle but some are symmetrically located around it, as can be seen in the upper left panel of Fig.~\ref{fig:Floquet_near_bif}, discussed in more detail below. None of the multipliers are outside the unit circle $|\rho|=1$, so the corresponding traveling wave solution is considered to be linearly stable. In the right panel, two additional real multipliers symmetric about the corresponding circle appear, with the larger one satisfying $\rho>1$. Hence the corresponding solution, pertaining to the
decreasing portion of the kinetic curve, is unstable.

\begin{figure}
\centerline{\psfig{figure=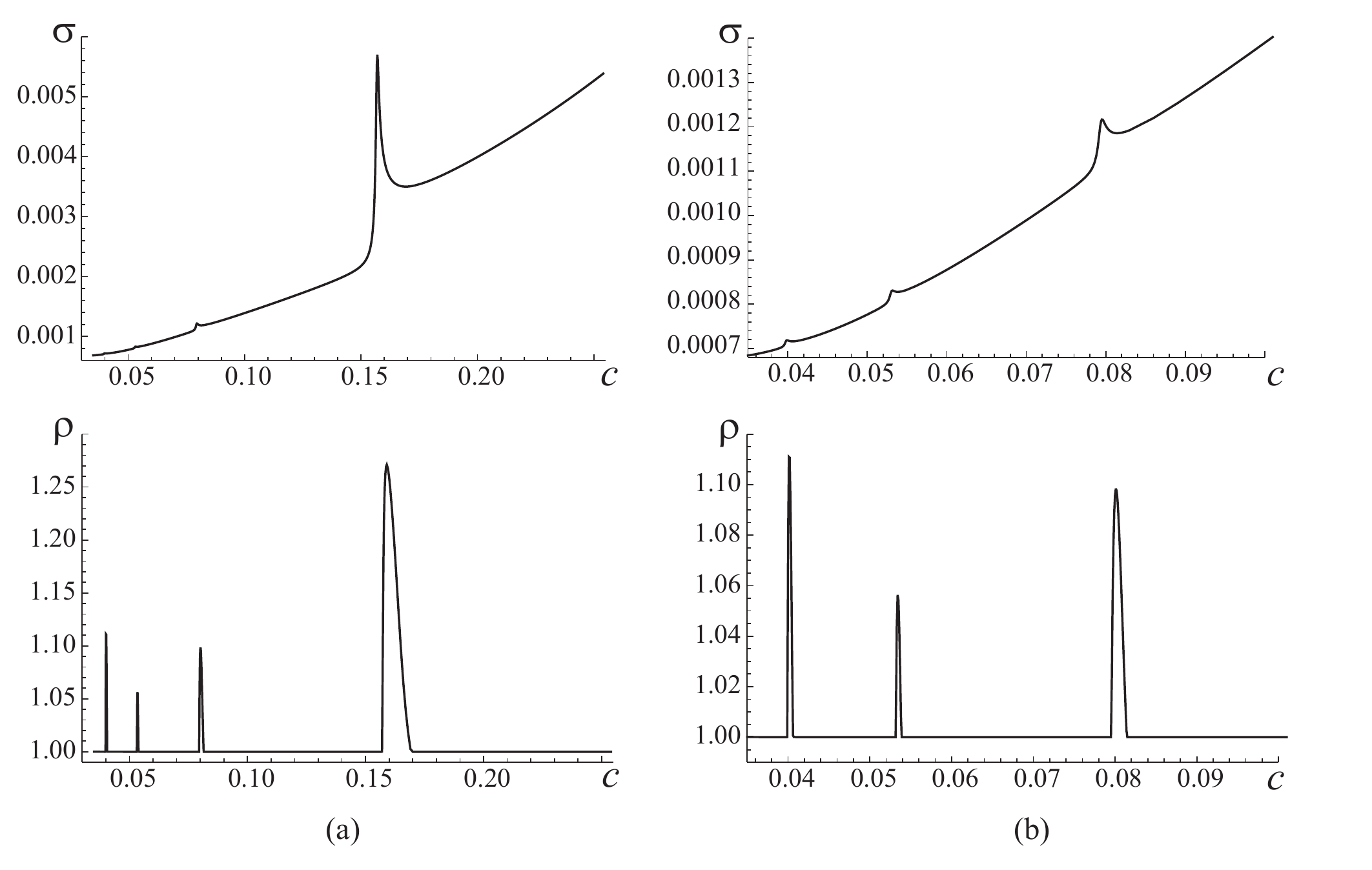, width=\textwidth}}
\caption{\footnotesize (a) Force $\sigma$ (upper panel) and the real Floquet multiplier $\rho$ (lower panel) with the maximum modulus for the traveling wave solutions with velocity $c$. (b) Zoomed-in version at smaller velocities. Here $\mu=1$ and $\gamma=0.01$.}
\label{fig:Floquet_gam0p01}
\end{figure}

\begin{figure}[h]
\centerline{\psfig{figure=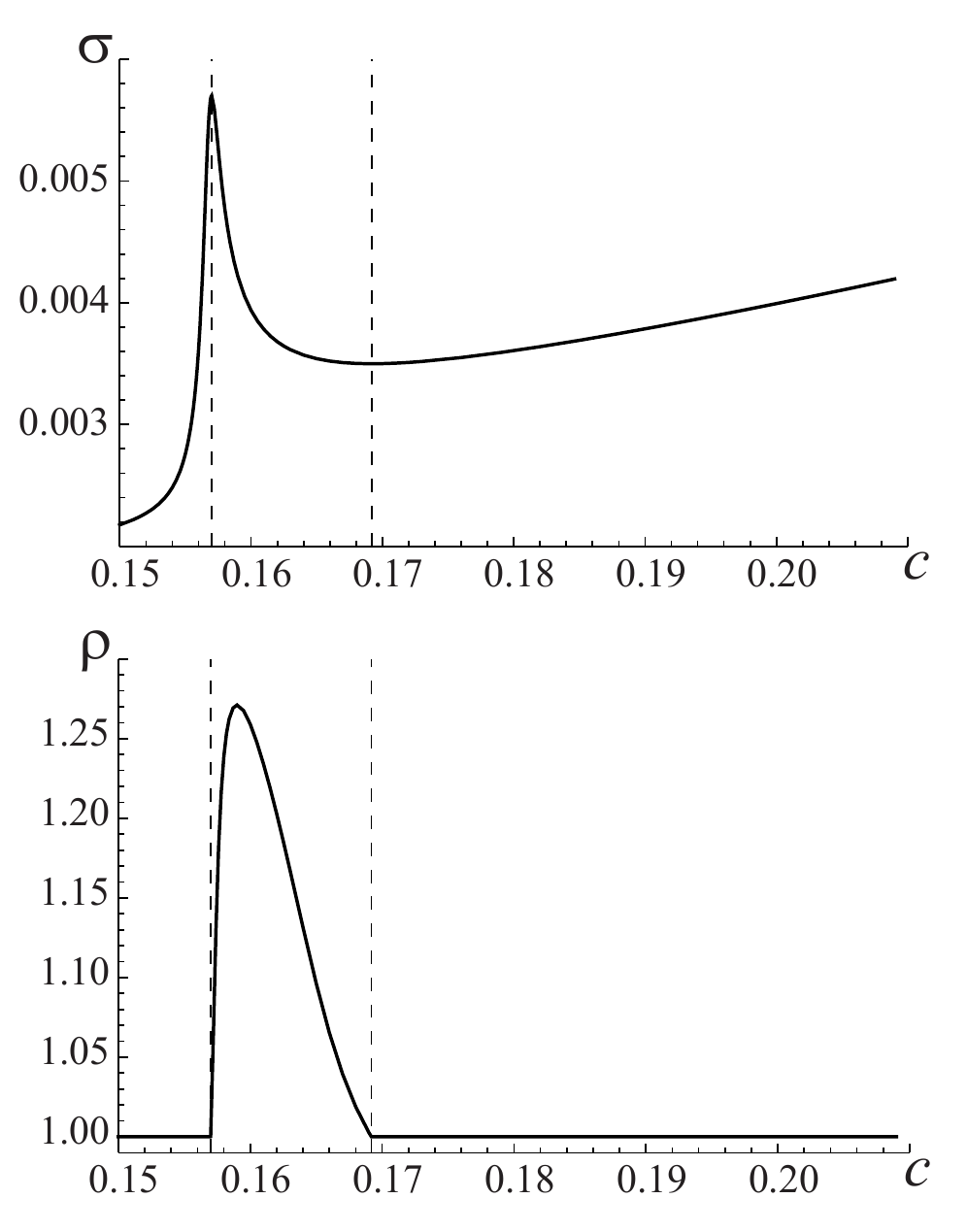, width=0.5\textwidth}}
\caption{\footnotesize Zoomed-in version of Fig.~\ref{fig:Floquet_gam0p01} around the first resonance. The dashed vertical lines mark the region where $\sigma'(c)<0$. Here $\mu=1$ and $\gamma=0.01$.}
\label{fig:Floquet_gam0p01_first_res}
\end{figure}

\begin{figure}[h]
\centerline{\psfig{figure=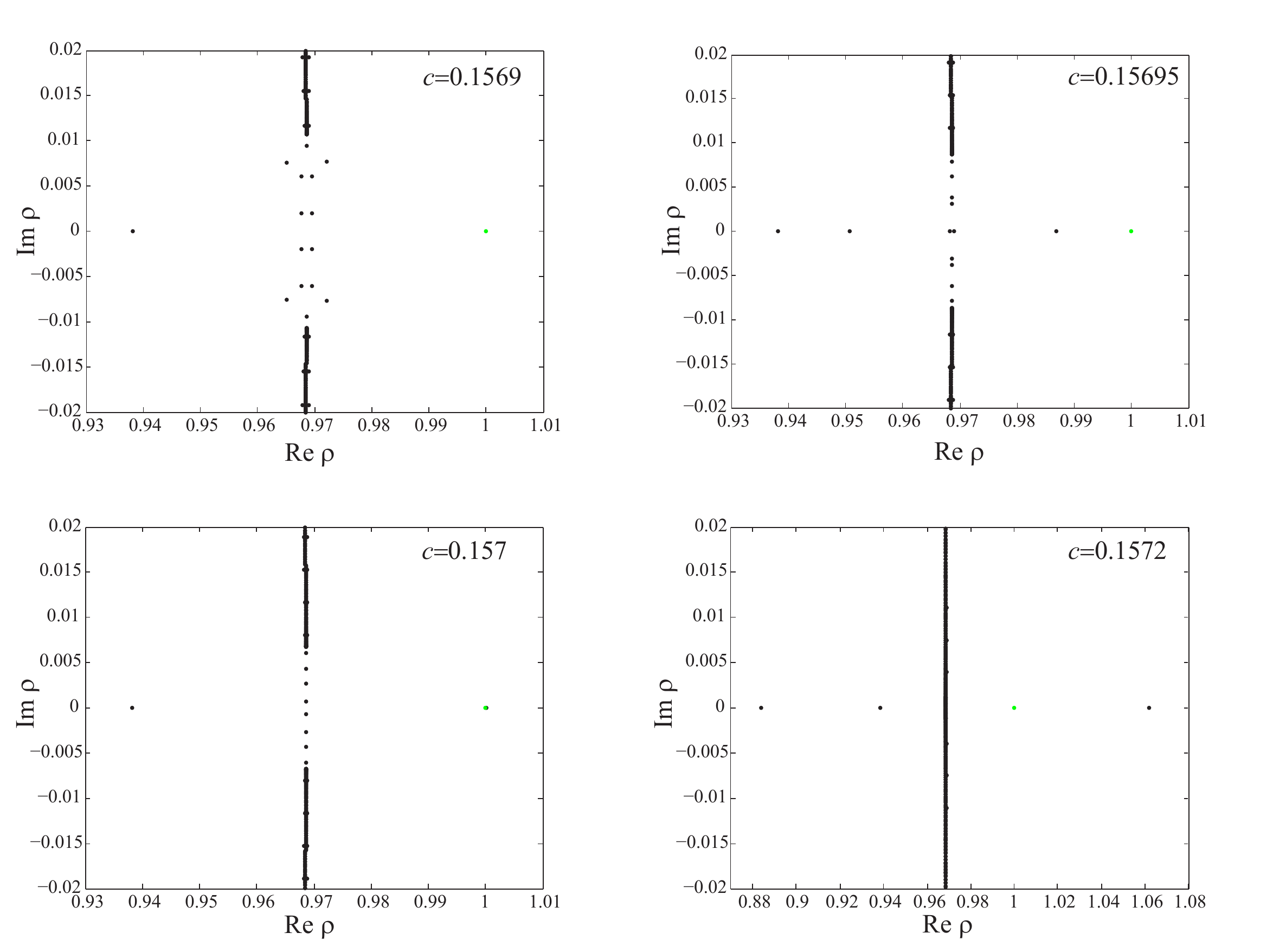,width=\textwidth}}
\caption{\footnotesize Floquet multipliers (dots) for traveling wave solutions at velocities near the stability threshold $c \approx 0.157$ { corresponding to a local maximum of $\sigma(c)$}. Here $\mu=1$, $\gamma=0.01$, { and the multiplier $\rho=1$ is marked by a green dot}.}
\label{fig:Floquet_near_bif}
\end{figure}

In Fig.~\ref{fig:Floquet_gam0p01}(a) we show the real multiplier $\rho$ that has the maximum modulus among all multipliers as a function of $c$ (lower panel) together with the corresponding $\sigma(c)$ plot (upper panel). Enlarged version of the same plots at smaller velocities is shown in Fig.~\ref{fig:Floquet_gam0p01}(b).
One can see that there are several intervals where $\rho>1$, implying instability. These intervals correspond to decreasing portions of the curve $\sigma(c)$. This can be clearly seen in Fig.~\ref{fig:Floquet_gam0p01_first_res}, which zooms in on the first resonance region.
In each case the transition between a (linearly) stable and unstable regimes occurs at velocities where $\sigma'(c)=0$. At these velocity values a real multiplier crosses the unit circle on the right.
It is important to highlight that this transition happens precisely at the extrema of
the kinetic curve, up to the numerical precision of our computations, i.e., the relevant
criterion appears to be {\it sharp}, suggesting a potential stability theorem, as we will further
discuss below.

\begin{figure}[h]
\centerline{\psfig{figure=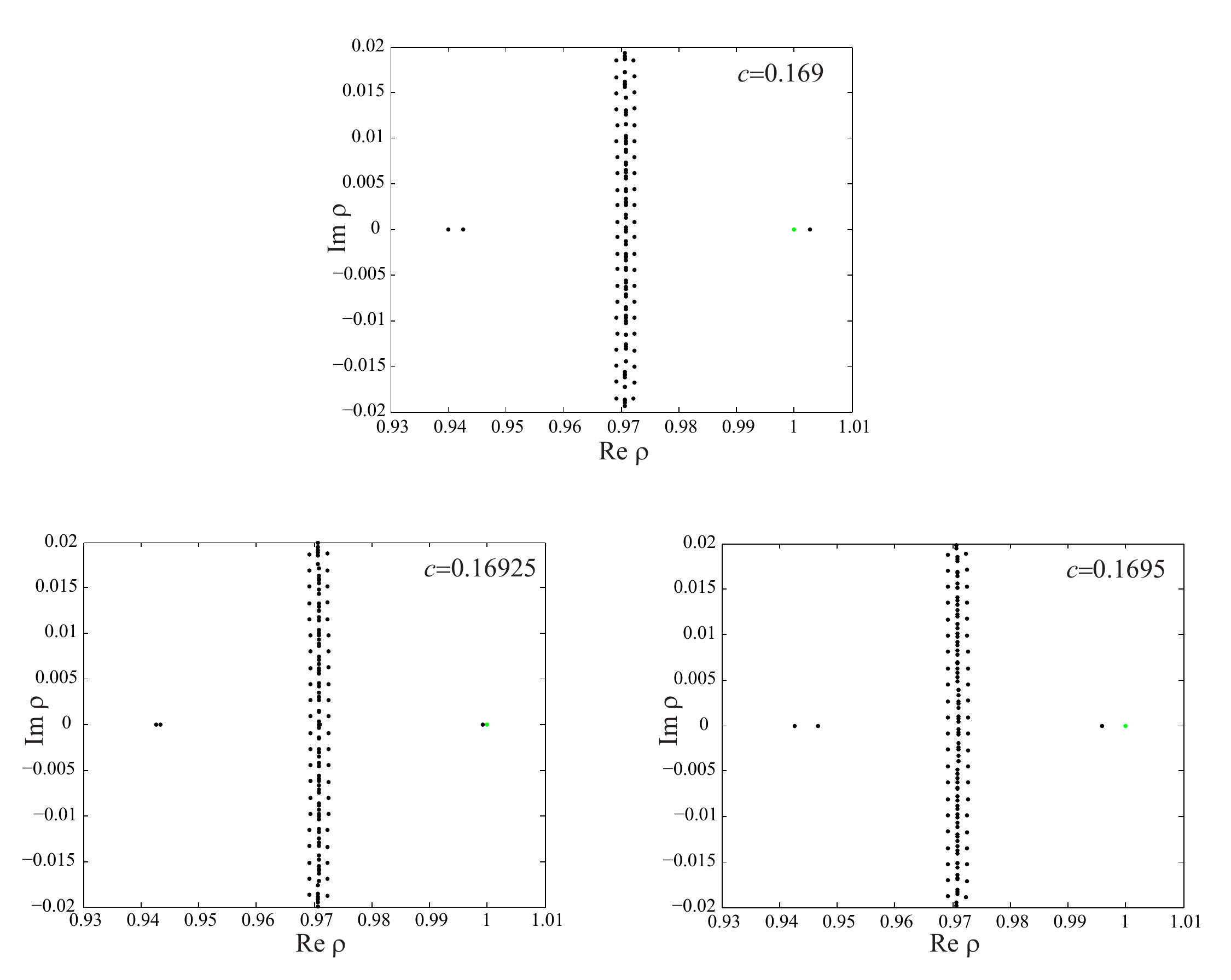,width=\textwidth}}
\caption{\footnotesize { Floquet multipliers (dots) for traveling wave solutions at velocities near the stability threshold $c \approx 0.16919$ corresponding to a local minimum of $\sigma(c)$. Here $\mu=1$, $\gamma=0.01$, and $\rho=1$ is marked by a green dot. The velocity is below the threshold in the upper panel and above it in the two lower panels.}}
\label{fig:Floquet_near_min}
\end{figure}

To further illustrate the mechanism for the instability, we show in Fig.~\ref{fig:Floquet_near_bif} the Floquet multipliers at velocities near the stability threshold $c \approx 0.157$, at which $\sigma'(c)$ changes sign from positive to negative.
One can see that at $c=0.1569$ (upper left panel) there are two real multipliers at $\rho_1=1$ and $\hat{\rho}_1=\exp(-\gamma/c)$, as well as various multipliers with nonzero imaginary part, some of which have exited the circle $|\rho|=\exp[-\gamma/(2c)]$ but still lie within the unit circle; thus
these all correspond to stable eigendirections. At $c=0.15695$, just below the stability threshold, the multipliers that were outside the circle have given rise to two additional pairs of real multipliers, $\rho_2$ and $\hat{\rho}_2=\exp(-\gamma/c)/\rho_2$ and $\rho_3$ and $\hat{\rho}_3=\exp(-\gamma/c)/\rho_3$ such that $\hat{\rho}_1<\hat{\rho}_2<\hat{\rho}_3<\exp[-\gamma/(2c)]<\rho_3<\rho_2<\rho_1=1$ (see the upper right panel). The multipliers $\rho_2$ and $\hat{\rho}_2$ then move to the right and to the left, respectively, and at velocity just below $0.157$, $\rho_2$ reaches the unit circle: $\rho_2=\rho_1=1$ and $\hat{\rho}_2=\hat{\rho}_1=\exp(-\gamma/c)$; see the lower left panel, where $c=0.157$ is slightly above the threshold value. As the velocity is further increased, the multipliers $\rho_2$ and $\hat{\rho}_2$ continue moving along the real axis, so that above the threshold, $\rho_2$ is outside the unit circle, as can be seen in the lower right panel, $c=0.1572$. The same mechanism, with a real exit multiplier appearing and merging with the unit one, was described in \cite{Braun00} in the discussion of the onset of the instability of fast kinks, which also takes place when $\sigma'(c)=0$.
{ Once the real multiplier $\rho_2>1$ corresponding to the unstable eigenmode reaches its maximum value (see Fig.~\ref{fig:Floquet_gam0p01_first_res}), it turns around and moves to the left along the real axis (while its counterpart $\hat{\rho}_2$ moves to the right) until it crosses the unit circle at $c \approx 0.16919$ corresponding to a local minimum of $\sigma(c)$. This is illustrated in Fig.~\ref{fig:Floquet_near_min}.}

\begin{figure}[h]
\centerline{\psfig{figure=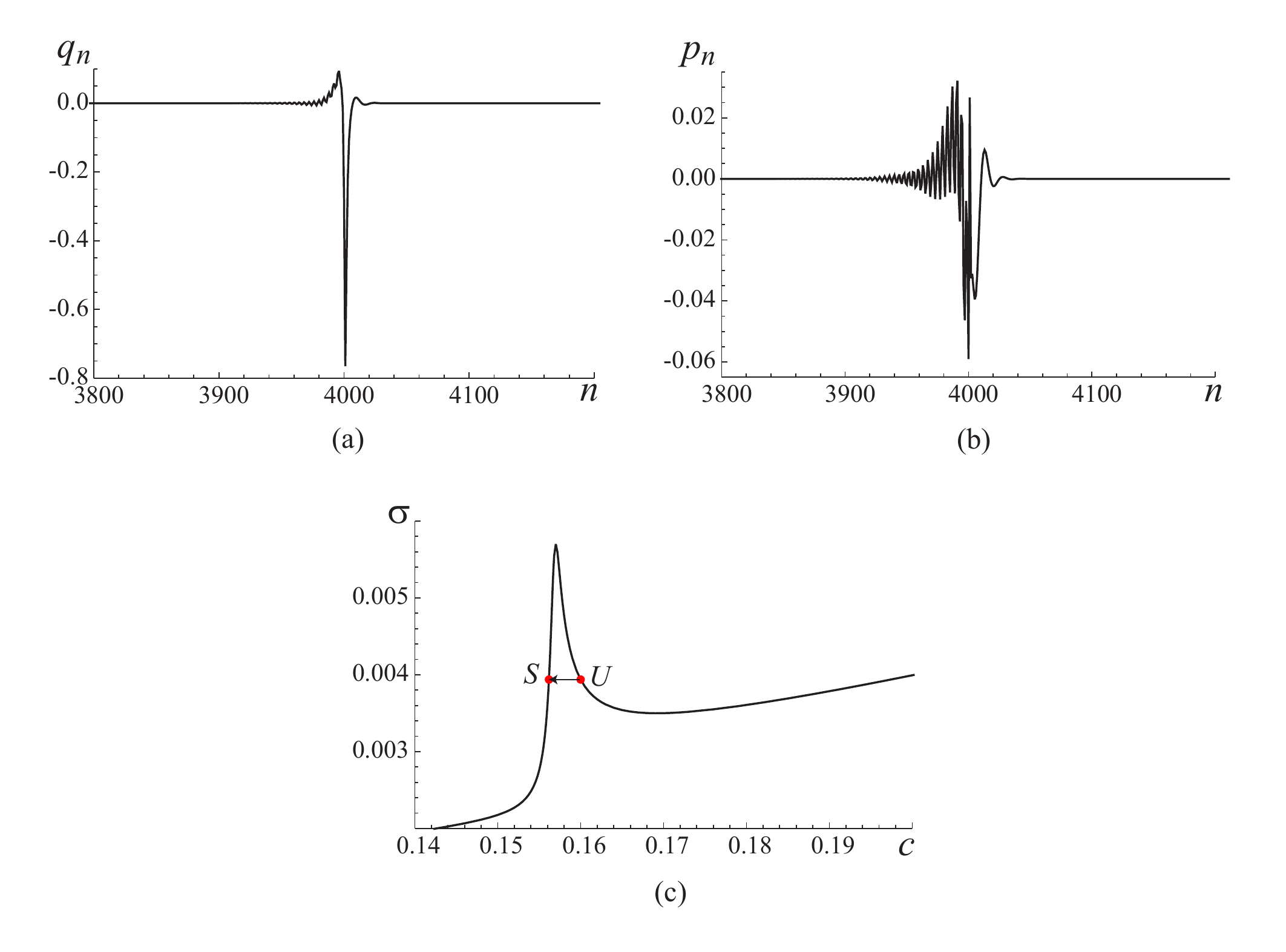,width=\textwidth}}
\caption{\footnotesize (a) displacement $q_n$ and (b) particle velocity $p_n$ for the unstable eigenmode at $c=0.16$ corresponding to $\rho=1.2591$ and (c) $\sigma(c)$ curve with the initial unstable (U) and final stable (S) states marked. The stable wave was obtained by solving \eqref{eq:dyn} with the initial conditions given by the unstable traveling wave solution perturbed along the unstable eigendirection (with perturbation amplitude $0.01$). It has velocity $c_f=0.1562$ and the same $\sigma=0.0039$ as the unstable wave. Here $\mu=1$ and $\gamma=0.01$.}
\label{fig:instab_c0p16gam0p01}
\end{figure}

\begin{figure}[h]
\centerline{\psfig{figure=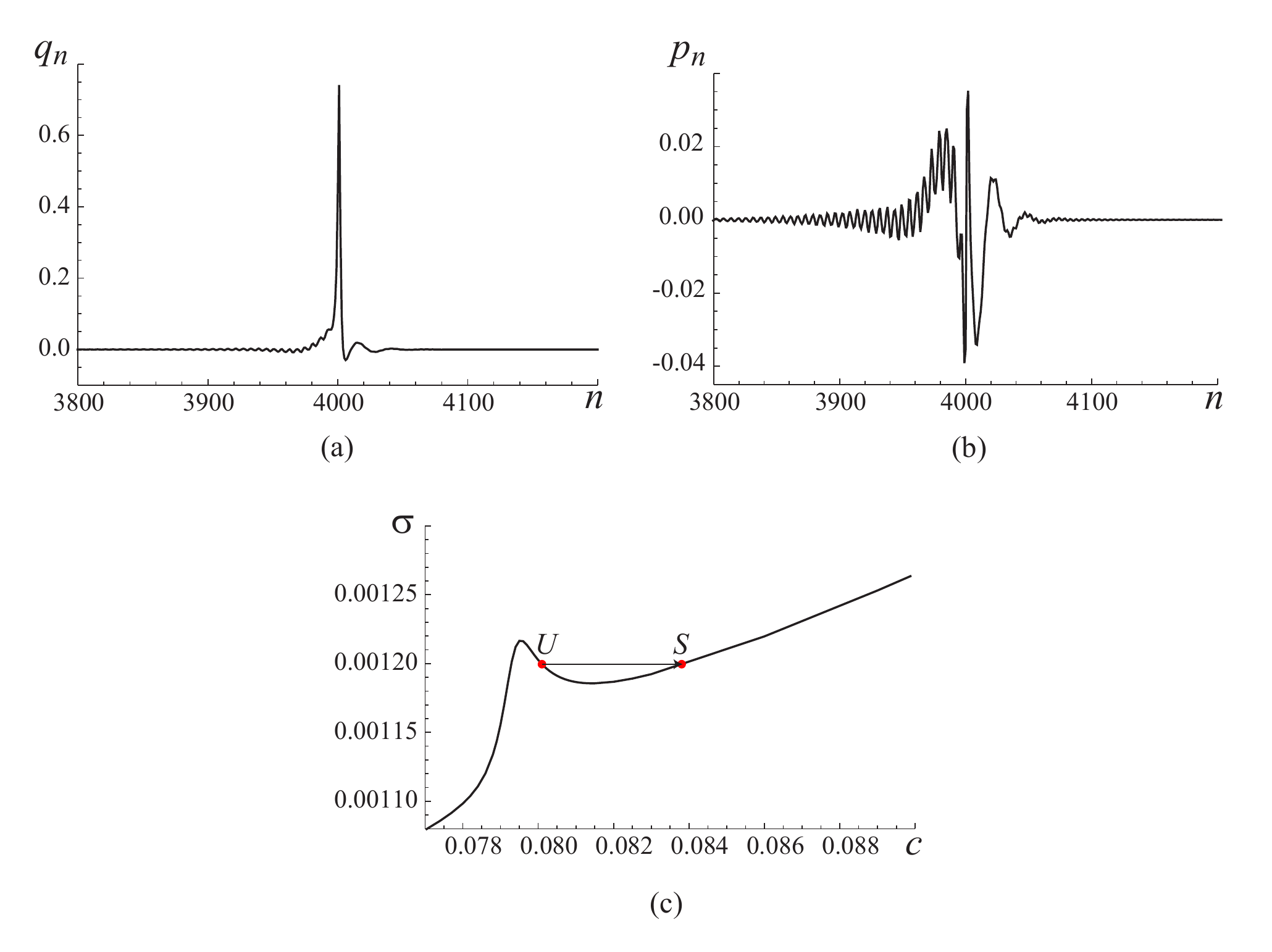,width=\textwidth}}
\caption{\footnotesize Same as the previous figure but now for $c=0.0801$ and for
  the unstable eigendirection with $\rho=1.0984$. As a result of the perturbation, in this
  case the evolution dynamics selects to move to the right stable branch portion corresponding to the
  higher speed $c_f=0.0838$ for the same value of $\sigma=0.0012$.}
\label{fig:instab_c0p0801gam0p01}
\end{figure}

To explore the dynamic consequences of the instability, we now consider
a typical example of an unstable evolution for $c=0.16$, a velocity value in an unstable region of $\sigma'(c)<0$,  with a real Floquet multiplier $\rho=1.2591$ for the corresponding traveling wave. The corresponding eigenmode is shown in Fig.~\ref{fig:instab_c0p16gam0p01}(a,b).
Perturbing the unstable wave along this eigendirection and solving \eqref{eq:dyn} with the resulting initial data, we find that the solution approaches a stable traveling wave with a lower velocity at the same $\sigma$, $c=0.1562$, as shown in Fig.~\ref{fig:instab_c0p16gam0p01}(c). If the sign of the perturbation is reversed, the solution instead approaches a faster stable wave, with $c=0.1974$. Similarly, when an unstable wave at $c=0.0801$ is perturbed along the corresponding eigenmode (see Fig.~\ref{fig:instab_c0p0801gam0p01}), the resulting solution approaches a stable one with larger velocity $c=0.0838$, while reversing the perturbation sign yields a slower stable solution with $c=0.0793$. Generally, we have found that
in an interval of bistability with the
two stable branches separated by the intermediate unstable one, it is possible to trigger the transition
to either one of the two stable branches, depending on the nature of the perturbation.

\subsection{Multivalued kinetic relation and instability at high velocities}
\label{sec:large_gamma}

In the previous section we provided an example of traveling waves becoming unstable due to the change in monotonicity of the kinetic curve
associated with small-velocity resonances. One expects the resonance peaks to disappear when damping is large enough. To illustrate this scenario and explore the other features of the kinetic curve, we now consider the case $\mu=1$, $\gamma=0.1$. This parameter regime was previously considered in \cite{Braun00}, where an iterative numerical method was used to compute traveling wave solutions as the force $\sigma$ was varied, and stability of the obtained solutions was investigated by computing the Floquet multipliers. In contrast to the fixed point approach employed here, the procedure in \cite{Braun00} yielded only stable solutions (dynamic attractors) at each $\sigma$.

A remarkable finding from \cite{Braun00} is the existence of a maximal force $\hat{\sigma}_1$ for the existence of $2\pi$-kink solutions, which takes place at a certain critical velocity $\hat{c}_1$. At $\hat{\sigma}_1$, the traveling wave becomes unstable, giving rise to creation of kink-antikink pairs in the tail of the primary kink at forces above this threshold. Using the Floquet analysis, the authors in \cite{Braun00} show that this instability takes place via the same mechanism that we described in Sec.~\ref{sec:small_gamma}, namely, a real multiplier crossing the unit circle on the right at $\sigma=\hat{\sigma}_1$. Using continuation in the damping constant, they find the critical force and velocity values for a range of $\gamma$ at two different values of $\mu$. However, the kinetic curve $\sigma(c)$ they compute (for $\gamma=0.1$ and $\mu=1$) terminates at the maximal force.

Here we revisit this parameter regime using our fixed point approach, with the goal to investigate what happens beyond the critical velocity $\hat{c}_1$. Since the higher damping leads to faster decay of tail oscillations, a smaller lattice size, $N=2000$, than used in our previous example is sufficient to accurately capture traveling wave solutions in the entire velocity range of their existence. To facilitate comparison with \cite{Braun00}, we use the same periodic topological charge-preserving boundary conditions \eqref{eq:per_BCs}.

\begin{figure}
\centerline{\psfig{figure=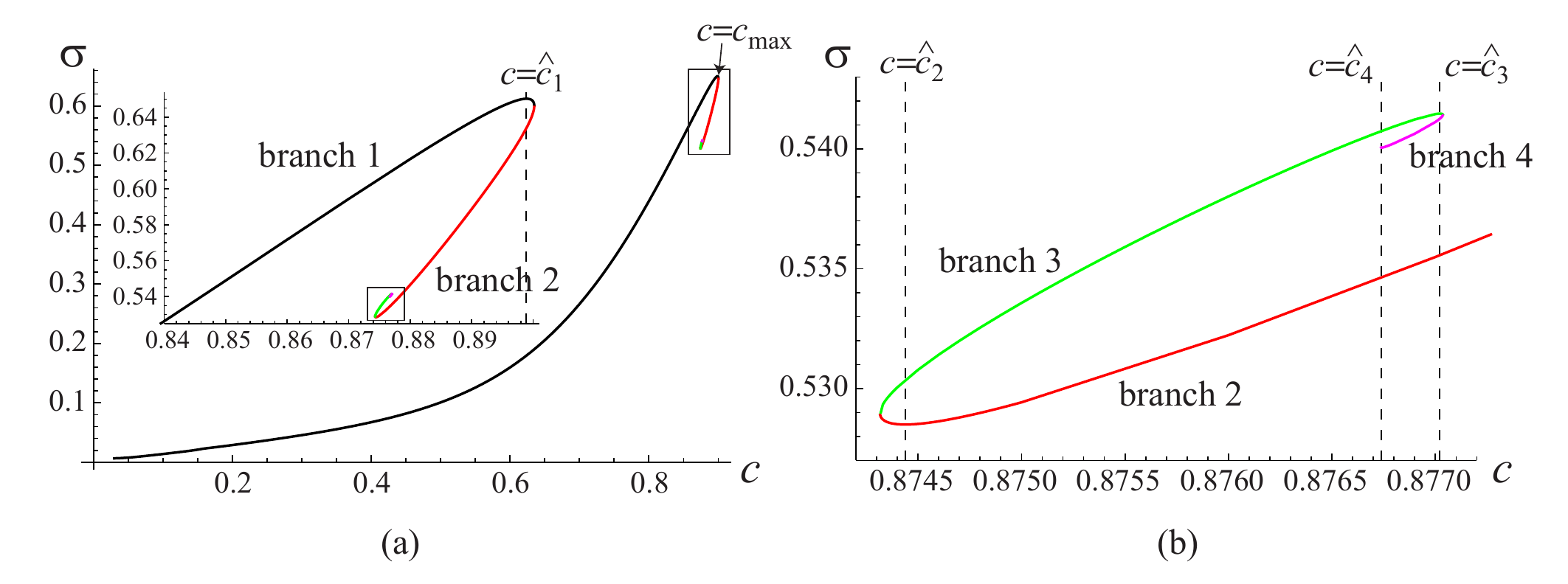,width=\textwidth}}
\caption{\footnotesize (a) The kinetic curve $\sigma(c)$. Here $\mu=1$ and $\gamma=0.1$. The inset zooms in on the multivalued part of the curve inside the rectangle. (b) Zoomed-in version showing the second, third and fourth branches inside the rectangle in the inset of (a).}
\label{fig:largeg_velocities}
\end{figure}

The resulting kinetic curve $\sigma(c)$, shown in Fig.~\ref{fig:largeg_velocities}, is in excellent agreement with the one computed in \cite{Braun00} for velocities below the critical point $\hat{c}_1=0.8989$, where $\sigma$ reaches its maximum value $\hat{\sigma}_1=0.65019$ (although our curve extends to smaller velocities). Unlike the small-damping case considered earlier, the curve exhibits no change in monotonicity near the resonance velocities, as can be seen in Fig.~\ref{fig:largeg_velocities}(a). Instead, in the vicinity of the first resonance velocity $c_1^*$ we observe a pair of real Floquet multipliers exiting the circle of radius $\exp[-\gamma/(2c)]$ within the velocity interval $[0.1569,0.1607]$, although in its excursion, the maximum value of $\rho$ is $0.74178$, occurring for $c=0.1583$. Hence, it never approaches a point of stability change that would necessitate $\rho=1$.

\begin{figure}
\centerline{\psfig{figure=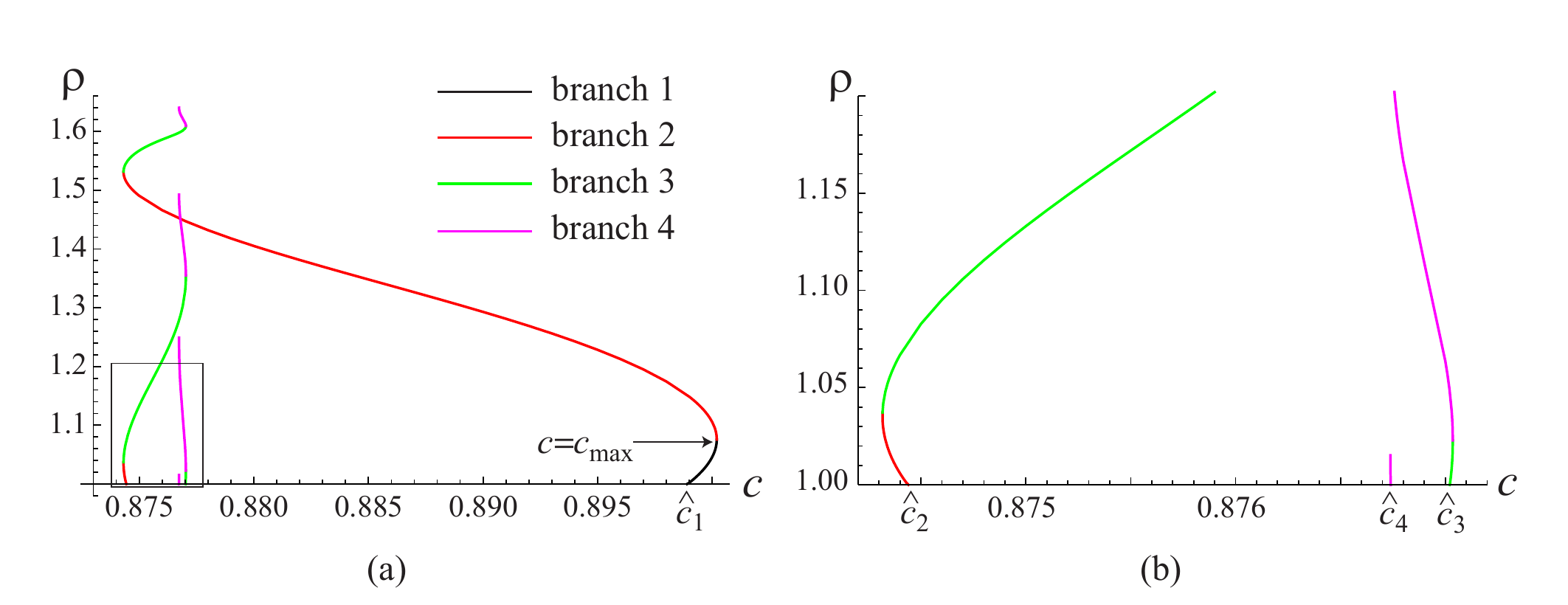,width=\textwidth}}
\caption{\footnotesize (a) Real Floquet multipliers $\rho>1$ in the velocity interval where kinetic relation is multivalued at $\mu=1$ and $\gamma=0.1$. (b) Enlarged view of the region inside the rectangle in (a). Along the $n$th branch, the number of unstable eigendirections changes from $n-1$ to $n$ at $c=\hat{c}_n$.}
\label{fig:largeg_floquetunstable}
\end{figure}

\begin{figure}
\centerline{\psfig{figure=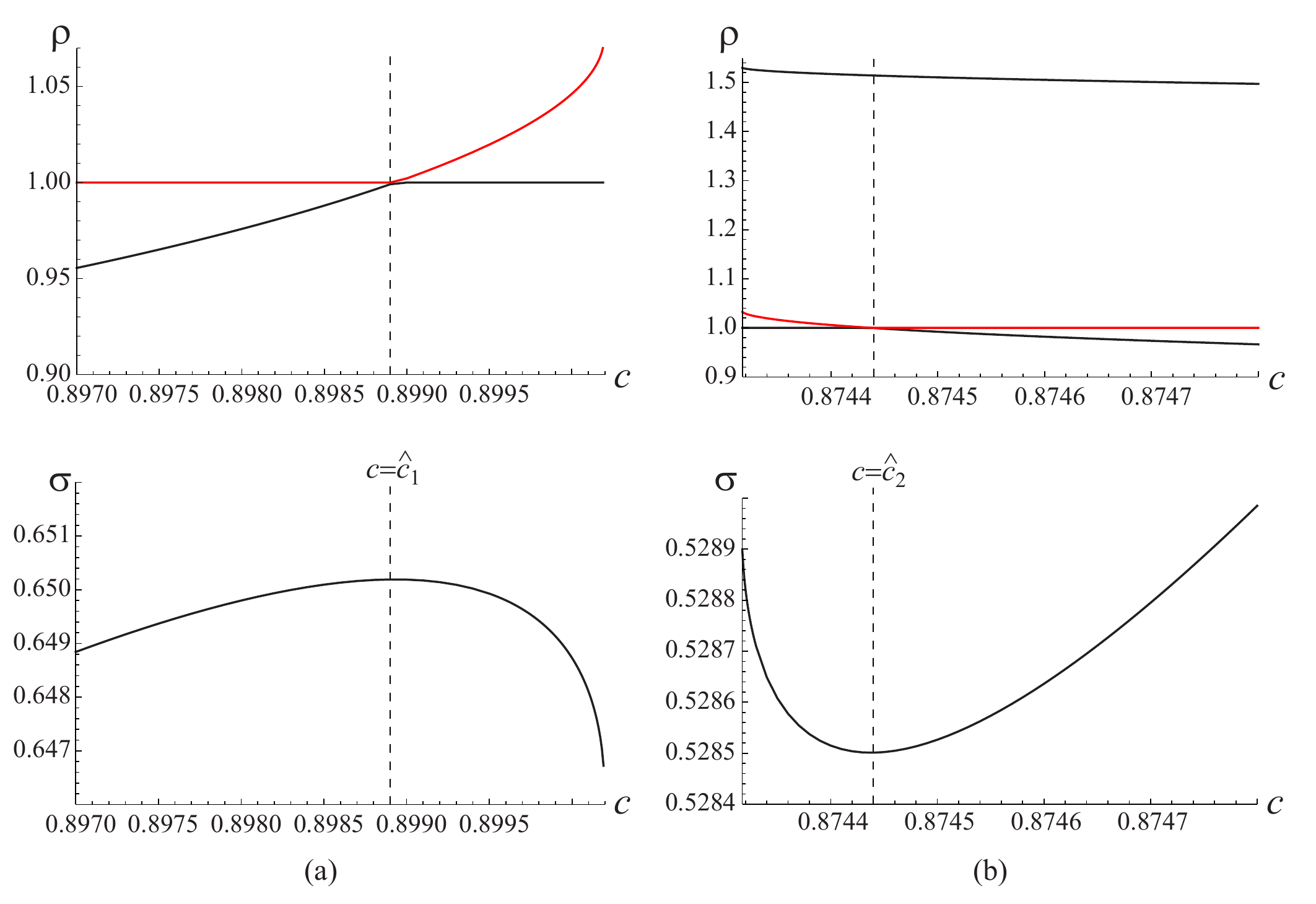,width=\textwidth}}
\caption{\footnotesize (a) Real Floquet multipliers $\rho>\exp(-\gamma/(2c))$ near the maximum of $\sigma(c)$ along the primary branch at $c=\hat{c}_1$ and the kinetic curve at the same region. Here $\mu=1$ and $\gamma=0.1$. (b) Same as (a) but for the second branch around the minimum of $\sigma(c)$ at $c=\hat{c}_2$. In both cases, the curve along which $\rho$  changes from $1$ to $\rho>1$ is marked by red.}
\label{fig:largeg_branch12}
\end{figure}

\begin{figure}
\centerline{\psfig{figure=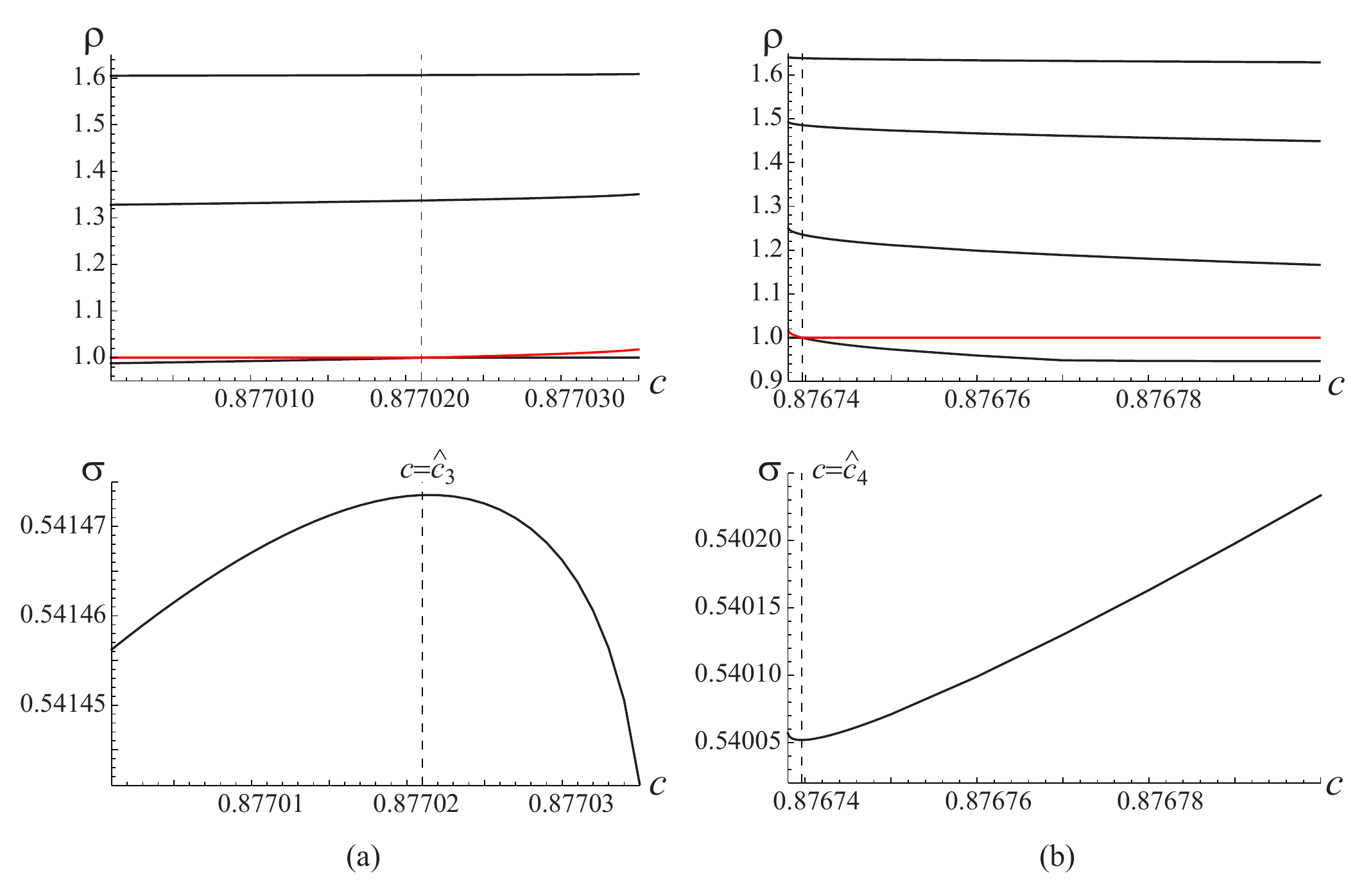,width=\textwidth}}
\caption{\footnotesize Same as in Figure \ref{fig:largeg_branch12} but near the maximum of $\sigma(c)$ along the third branch at $c=\hat{c}_3$ in (a) and near the minimum of $\sigma(c)$ along the fourth branch at $c=\hat{c}_4$ in (b).}
\label{fig:largeg_branch34}
\end{figure}

At $c=\hat{c}_1$ a real Floquet multiplier crosses the unit circle on the right, as shown in Fig.~\ref{fig:largeg_floquetunstable}(a) and Fig.~\ref{fig:largeg_branch12}(a), and as we continue the kinetic curve beyond this critical velocity value (and thus beyond the portion of the curve computed in \cite{Braun00}), with the force $\sigma$ now decreasing, the corresponding traveling waves are unstable, with a real Floquet multiplier $\rho>1$. Remarkably, we found that there is not only a maximal force but also a \emph{maximal speed} $c=c_\text{max}=0.9002$ for existence of traveling wave solutions; see Fig.~\ref{fig:largeg_velocities}. { To compute the new branches, continuation in $\sigma$ was used in the vicinity of the turning points, while the remaining parts were computed using continuation in $c$.}
At this point, the kinetic curve turns around, giving rise to another branch of the kinetic relation $\sigma(c)$, which is thus \emph{multivalued}. As the kinetic curve continues beyond the turning point, the traveling waves remain unstable, as shown in Fig.~\ref{fig:largeg_floquetunstable}. More specifically, the new (second) branch of the kinetic relation features exponentially unstable solutions with a single unstable eigendirection for $\hat{c}_2\leq c\leq c_\text{max}$, where $\hat{c}_2=0.87444$ corresponds to a local minimum of $\sigma(c)$. We note that this instability along an increasing portion of the curve does not contradict the stability criterion, which states that $\sigma'(c)<0$ is sufficient but not necessary for instability. For $c<\hat{c}_2$, there is an additional portion of the kinetic curve with $\sigma'(c)<0$ and, again in accordance with the criterion, this leads to the emergence of a \emph{second} real Floquet multiplier $\rho>1$ and an additional unstable eigendirection, as shown in Fig.~\ref{fig:largeg_floquetunstable} and Fig.~\ref{fig:largeg_branch12}(b). This segment terminates at $c=0.87432$, giving rise to yet another branch of the kinetic relation. This third branch, whose stability properties are illustrated in Fig.~\ref{fig:largeg_floquetunstable} and Fig.~\ref{fig:largeg_branch34}(a), possesses a local maximum of the kinetic curve at $c=\hat{c}_3=0.87021$ so that for $c>\hat{c}_3$ a \emph{third} real multiplier with $\rho>1$ emerges. This branch terminates at $c=0.877035$, giving rise to a fourth branch with a local minimum in the kinetic curve at $c=\hat{c}_4=0.87674$ (see Fig.~\ref{fig:largeg_floquetunstable} and Fig.~\ref{fig:largeg_branch34}(b)), so that a \emph{fourth} unstable eigendirection arises for $c<\hat{c}_4$; the branch ends at $c=0.87638$. This branching phenomenon seems to occur {\em ad infinitum}, with the length of the kinetic curve along each new branch smaller than the previous one. We conjecture that this self-similar pattern
continues in a spiral fashion, with the number of unstable eigendirections progressively increasing.

{ To further understand the spiraling nature of the kinetic curve, we consider the evolution of the solution profiles along the spiral. In Fig.~\ref{fig:spiral_solns} we show the displacements $u_n(0)=\phi(n)$ for the solutions along the four branches at the velocities $\hat{c}_i$, $i=1,\dots,4$, where $\sigma(c)$ reaches an extremum (maximum along the first and third branches and minimum along the second and fourth).
\begin{figure}
\centerline{\psfig{figure=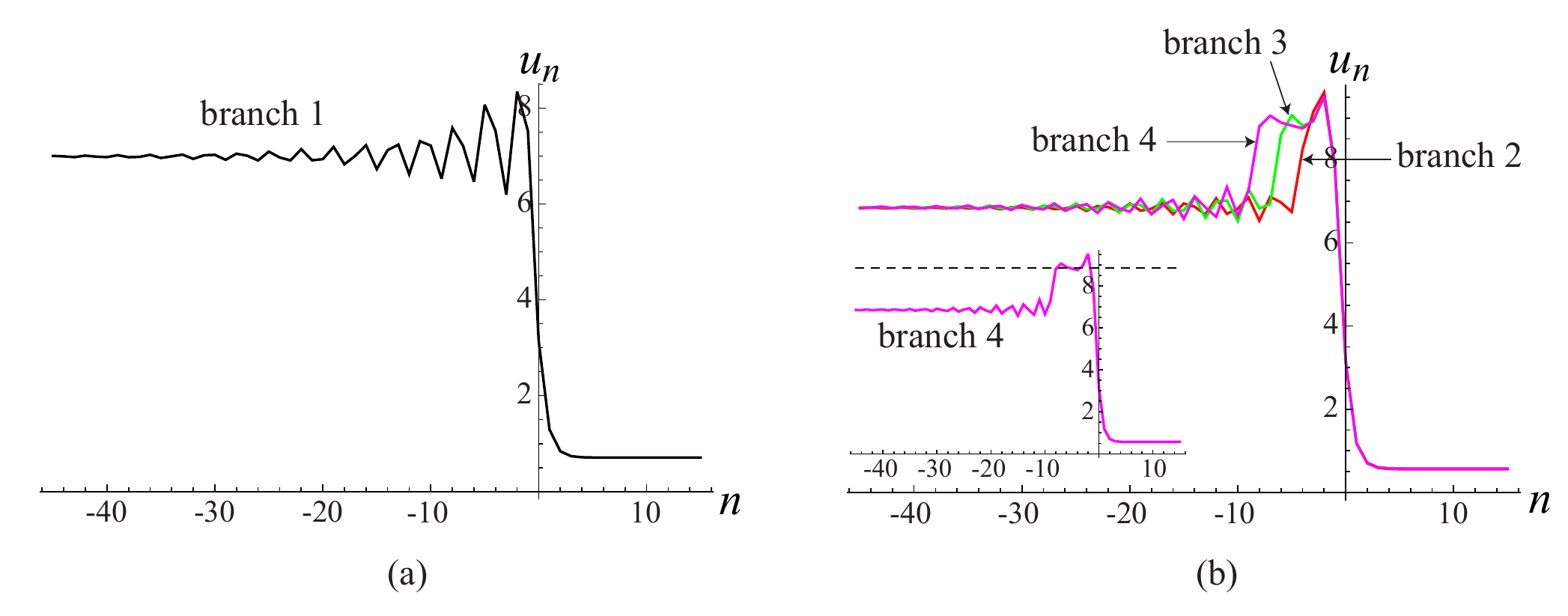,width=\textwidth}}
\caption{\footnotesize { Displacement profiles $u_n(0)=\phi(n)$ (a) along the first branch at $c=\hat{c}_1$; (b) along the second, third and fourth branches, at $c=\hat{c}_2$, $c=\hat{c}_3$ and $c=\hat{c}_4$, respectively. Inset in (b) shows the profile along the fourth branch with the equilibrium state $3\pi-\text{arcsin}(\sigma)$ marked by the dashed line for the corresponding $\sigma \approx 0.54$. Here $\mu=1$ and $\gamma=0.1$.}}
\label{fig:spiral_solns}
\end{figure}
One can see that the solution at $c=\hat{c}_1$ along the first branch (panel (a)) features a decaying oscillation behind the front around the equilibrium state $\text{arcsin}(\sigma)+2\pi$, with $\sigma=\hat{\sigma}_1$. As we go further along the spiral, the structure of the solutions changes. As shown in panel (b), a portion of the chain behind the front oscillates around a higher (unstable) equilibrium state, $3\pi-\text{arcsin}(\sigma)$, before settling into the stable state $\text{arcsin}(\sigma)+2\pi$ for the corresponding $\sigma$. The size of this portion increases as we trace the curve toward the spiral center. This suggests that in the limit the solution approaches a trajectory that connects the unstable state to the stable equilibrium $\text{arcsin}(\sigma)$.

The spiraling phenomenon we just described is a discrete analog of the one discussed in \cite{Brown94,Maksimov96,vdBerg03} for a related continuum model of a long Josephson junction governed by the damped driven (singularly perturbed) sine-Gordon equation
\beq
u_{xx}-u_{tt}-\sin u=\gamma u_t-\beta u_{xxt}-\sigma,
\label{eq:ddSG}
\eeq
where $u(x,t)$ is the quantum mechanical phase difference, $\sigma$ is the bias parameter satisfying $0<\sigma<1$, and both viscosity coefficients $\gamma$ and $\beta$ are assumed to be positive (note that at $\beta=0$ \eqref{eq:ddSG} is the continuum analog of \eqref{eq:dyn} with $\mu=1$). In particular, numerical results in \cite{Brown94,Maksimov96} show a spiral centered at $c=1$ in the relation between $\sigma$ and $c$ for traveling wave solutions $u(x,t)=\phi(x-ct)$, $c>0$, connecting the equilibrium states $\text{arcsin}(\sigma)$ and $\text{arcsin}(\sigma)+2\pi$ (considered to be homoclinic orbits in \cite{vdBerg03} since the two limiting states correspond to the same angle). As the spiral center is approached, the solutions are close to the unstable equilibrium state $3\pi-\text{arcsin}(\sigma)$ along progressively longer spatial intervals. This is similar to what we show in Fig.~\ref{fig:spiral_solns}(b), although there are no oscillations in the continuum case. Numerical stability analysis shows that none of these solutions are stable beyond the first extremal point of the kinetic curve, and the number of unstable eigenmodes progressively increases as one passes the subsequent extrema, again in agreement with our observations. At the center of the spiral,
the traveling wave equation acquires additional symmetries that lead to the existence of a monotone \emph{heteroclinic} trajectory connecting the unstable and stable states, as  proved in \cite{vdBerg03}. The more general conditions for breaking of a heteroclinic orbit into homoclinic ones is discussed in the earlier work of~\cite{Bykov00}. Proving the analogous result for the discrete case is an interesting problem to consider in  future work. It should also be noted that despite the apparent similarities of the spiraling mechanism in the discrete and continuum settings, some details are different. In addition to the absence of oscillations, in the continuum problem the turning points in $\sigma$ and $c$ coincide \cite{Brown94}, while this is not generally the case for the discrete FK problem, as our results demonstrate. Furthermore, the spiral center is always at $c=1$ in the continuum case \cite{vdBerg03}, while our results yield $c \approx 0.877$. Further studies are necessary to elucidate the passage from a strongly discrete setting to the continuum limit in this context.}

\section{Conclusions, discussion and future work}
\label{sec:conclusions}

In this work, we have revisited the topic of traveling waves in
damped and driven Frenkel-Kontorova lattices and have provided a systematic
numerical perspective on how to compute such solutions numerically, via a {
fixed point method with Newton-Raphson iterations} that enables the identification
of both stable and unstable segments of the kinetic curve $\sigma=\sigma(c)$ of the force as a function
of speed. This, in turn, allowed a definitive computation of
the Floquet multipliers associated with the solution considered as a periodic orbit
(modulo shifts). We have argued both here and in our earlier works~\cite{jxu1,jxu2}
that this is a beneficial way of examining traveling wave solutions on a lattice
as it helps to understand their stability and hence provides an informed
view on their dynamics.

While the Hamiltonian case of drive in the absence of damping remains somewhat
elusive for our computational approach due to non-decaying quasiperiodic tail oscillations,
we have considered the cases of both weaker and stronger damping. In the former case, some of the resonance velocities, corresponding
to the change in the number of modes emitted by the traveling wave in the undamped lattice, are still prominent, but only lead to a local maximum of the kinetic curve $\sigma=\sigma(c)$ and a bistable region in its vicinity. Most notably,
we observe in as definitive a way as our numerical precision allows that a change
of stability arises when the kinetic curve changes from increasing to decreasing.
For the decreasing portions of the curve, we always observe an unstable eigendirection
associated with a real Floquet multiplier larger than unity.

When the damping is strong enough, small-velocity resonances no longer affect the stability of the traveling waves and the corresponding
monotonicity of the kinetic curve. However, the curve becomes non-monotone in the large-velocity regime. In fact, as we have shown, it also becomes multivalued and features a maximal velocity at a turning point beyond which there are no traveling waves, an observation that to our knowledge has been missed in the earlier studies { of the discrete problem.} In this case, too, the traveling waves become unstable once the primary branch reaches its peak. The waves remain unstable along the secondary branches. Importantly though, every time a branch goes through an extremum point where $\sigma'(c)=0$, crossing it towards the decreasing portion of the curve, an additional real multiplier crosses the unit circle on the right, and an additional unstable eigendirection appears.

The realization that $\sigma'(c)<0$ leads to a dynamical instability
is rather remarkable in its own right, in our view. Firstly, this is so because the
sharpness of the relevant criterion suggests that it is highly likely to be associated
with a theorem.  Secondly, the stability criterion is the
same as the one that emerges from the Hamiltonian analogue of the problem, e.g.,
in~\cite{EW77,V10}. The fact that the Hamiltonian and dissipative variants of the problem
feature the same stability criterion may seem surprising. Indeed, one may think that
the Hamiltonian criterion provides only an approximate, rather than exact, separatrix
between stability and instability in the dissipative case. However, the
discernibility of the stable and unstable multipliers in Fig.~\ref{fig:Floquet_examples}
and the numerical sharpness of the criterion in Fig.~\ref{fig:Floquet_gam0p01_first_res}
suggest that this is not the case here. Thirdly, this criterion is similar in spirit
with the criteria that we have established in our recent works in~\cite{jxu1,jxu2}
(and with the classic work of~\cite{Friesecke03}). We have, in fact, attempted
to extract such a ``theorem'' following the pattern of the corresponding proofs
in~\cite{Friesecke03,jxu1,jxu2}. However, this attempt encounters two significant
stumbling blocks. The first one is technical: the solutions considered here feature
slow decay to a potentially nontrivial asymptotic state. Hence, the spaces used
and the inner product definitions therein need to be suitably adapted accordingly
to follow the proof. However, the second problem is conceptual and,
arguably, deeper: the way
that the proof proceeds involves the realization that in the Hamiltonian problem,
the Floquet multiplier with $\rho=1$ bears two eigendirections: an eigenvector $\partial_{\xi} \phi$
and a generalized eigenvector with $\partial_c \phi$, where $\phi(\xi)$ is the traveling wave.
Then, the stability criterion stems from
the {\it symplectic orthogonality} of an additional eigenvalue ``colliding'' with
$\rho=1$ with these eigendirections and, in fact, most notably with the generalized
eigenvector. Yet, the key realization for our setting is that such a generalized eigendirection
is {\it eliminated} in the presence of dissipation: rather, as discussed in our numerical
results, the relevant multiplier now lies at $e^{-\gamma/c}$. Hence, the very core
of the calculation (see, for example, the line between Eqs.~(5) and~(6) in~\cite{jxu1}) is no
longer immediately valid.

Despite the above discussion, we find that the ``Hamiltonian criterion'', namely, that $\sigma'(c)<0$
leads to instability, is definitively verified in our numerical computations. This poses a nontrivial
challenge from the point of view of analysis to establish the stability criterion (or perhaps disprove it, meaning that it
is not exact but rather only approximate). The fact that the relevant multiplier
circle is at $e^{-\gamma/(2 c)}$ and the complementary multiplier is $e^{-\gamma/c}$ suggests
that perhaps in some suitably weighted space the situation can be reverted to the Hamiltonian frame
and hence the proof of the criterion can be reconstructed in a way reminiscent of the original
one. Another substantial challenge concerns the systematic identification of solutions
at the undamped limit, complicated by the presence of quasiperiodic tail oscillations in the small-velocity regime.
In particular, it would be interesting to see whether the kinetic relation retains its multivalued nature and bifurcation structure at large velocities in this limit.
These topics are currently under consideration and will be reported in the future
publications.

\vspace{5mm}

{\it Acknowledgements.} PGK gratefully acknowledges informative discussions with
A.R. Champneys and with H. Susanto (who also pointed out Ref.~\cite{vdBerg03}))
on this topic, as well as the hospitality of the Mathematical
Institute of the University of Oxford and the support of the Leverhulme Trust
during the final stages of this work. This
material is based upon work supported by the US National Science Foundation under Grant
DMS-1809074 (AV and PGK), MAT2016-79866-R project (AEI/FEDER, UE) (JCM) and NSFC (Grant No. 11801191) (HX).

\end{document}